\documentclass[letterpaper,11pt]{article} 

\usepackage[table]{xcolor}
\usepackage[T1]{fontenc}
\usepackage[english]{babel}
\usepackage[margin=3cm]{geometry}
\usepackage{amsmath,amsthm,amssymb,amsfonts,amstext,mathtools,mathrsfs}
\usepackage{xspace}
\usepackage{type1cm}
\usepackage{graphicx,epsfig,wrapfig,rotating}
\usepackage{color}
\usepackage{array,booktabs,hhline,multirow,multicol}
\usepackage{tabularx,pbox,makecell,caption,subcaption}
\usepackage{threeparttable,adjustbox}
\usepackage{enumerate,enumitem}
\usepackage{changepage,lipsum}
\usepackage{comment}
\usepackage{authblk}
\usepackage{natbib}
\usepackage{sgame}
\usepackage{apptools}
\usepackage{setspace}
\usepackage{hyperref}

\AtAppendix{\counterwithin{table}{section}}
\AtAppendix{\counterwithin{example}{section}}
\hypersetup{
colorlinks   = true, 
urlcolor     = blue, 
linkcolor    = blue, 
citecolor   = blue,   
hyperindex=true
}

\newtheorem{definition}{Definition}

\newtheorem{theorem}{Theorem}
\newtheorem{proposition}{Proposition}
\newtheorem{lemma}{Lemma}
\newtheorem{example}{Example}
\newtheorem{corollary}{Corollary}

\def \C{\mathcal{C}}

\def \I{\mathcal{I}}

\def \S{\mathcal{S}}

\def \M{\mathcal{M}}

\def \T{\mathcal{T}}

\title{Verifiable affirmative action in centralized school admissions\thanks{We thank Orhan Ayg\"{u}n and Bertan Turhan for comments. All errors are ours. 
	}
}

\author{Xinquan Hu\thanks{School of Business, Xiangtan University. Email: xinquan.hu.eco@gmail.com.} \qquad Jun Zhang\thanks{Institute for Social and Economic Research, Nanjing Audit University. Email: zhangjun404@gmail.com.}
}

\date{June 4, 2026}

\begin{document}

\maketitle	
\thispagestyle{empty}
\setcounter{page}{0}

\begin{abstract}\label{abstract}
	Governments increasingly operate centralized, algorithm-run admission clearinghouses that implement affirmative action through reserve systems. To sustain public trust, many such clearinghouses disclose category-specific cutoffs, but cutoffs need not allow participants to verify whether reserved and open seats are correctly assigned. We formulate cutoff-based verifiability as a governance constraint on the clearinghouse: each participant must be able to verify her assigned school and seat type using only her own score and the public cutoffs, under two intuitive verification protocols. In a controlled school choice model with multiple reserve categories, we characterize mechanisms that are individually rational, strategy-proof, and verifiable. The characterization identifies deferred acceptance mechanisms induced by two choice rules. We recommend one rule that assigns reserved seats only when a student cannot secure an open seat on merit, so that every reserved-seat assignment reflects genuine affirmative action. The results explain mechanism choices across China's high school admission systems and provide design guidance for affirmative action systems in Brazil and India.
\end{abstract}

\bigskip



\thispagestyle{empty}
\setcounter{page}{0}


\newpage

\section{Introduction}\label{section:introduction}

In 2025, a family in Shenzhen, China, reportedly discovered that a reserved high-school seat that they believed should have been assigned to their daughter had instead been given to an ineligible student from the same middle school, prompting public calls for greater transparency.\footnote{The incident happened in Shenzhen's 2025 high school admissions. See the \href{https://www.163.com/dy/article/K7GEE6IE0536BKXN.html}{news report} (in Chinese).} The episode reflects a broader challenge in the operation of modern admission clearinghouses. In China and many other countries, governments implement affirmative action policies in public school and university admissions through reserve systems that set aside quotas for targeted groups. Because the allocation is executed by a procedure the public cannot directly observe, these clearinghouses sustain trust by disclosing category-specific admission cutoffs and inviting participants to check that reserved seats are correctly assigned. This paper studies how to design such a clearinghouse so that this check is conclusive: every participant should be able to confirm both her assigned school and her seat type using only her own score and the disclosed cutoffs. We formalize this property, which we call \emph{verifiability}, as a transparency requirement on the design of the clearinghouse, and we characterize which allocation mechanisms satisfy it together with individual rationality and strategy-proofness.

Public disclosure of this kind is a common governance device in reserve-based affirmative action. In Brazil, the government discloses category-specific cutoff scores for public university admissions through the nationwide SISU platform, and some universities additionally publish the profiles of admitted students \citep{aygun2021college,francis-tan2024affirmative}. In India, affirmative action in public university admissions and government recruitment is similarly accompanied by disclosure of category-specific cutoffs and the names of admitted applicants \citep{aygun2017largescale,sonmez2022affirmative}. In China's high school admissions, policymakers must disclose cutoff scores for each seat category at each high school, and some cities further publish the profiles of students admitted through reserve quotas. These disclosure practices make participants' ability to verify outcomes a first-order design consideration. We conduct our analysis within the standard controlled school choice model (Section~\ref{section:model}), in which students have multiple types and each school has type-specific reserve quotas, so that the framework applies to any admission clearinghouse where seat-type assignments must be verifiable.

Verifiability arises as a substantive design constraint specifically because affirmative action creates multiple seat categories within each school. Seat-type assignments record how reserve quotas are used: a student assigned to a reserved seat is a beneficiary of affirmative action, so verifying seat types is equivalent to verifying that the policy has been applied correctly in her case. If all seats were identical, students could verify their assigned schools by comparing their scores against publicly disclosed school-level cutoffs, a well-known property of stable mechanisms. However, when schools reserve seats for specific groups, each student faces two relevant cutoffs at any given school: one for open seats and one for her group's reserved seats. When her score meets both cutoffs, her seat type is not determined by the cutoffs alone, and additional rules are required. Without a publicly known and consistent rule for resolving this ambiguity, students cannot consistently verify their seat-type assignments from disclosed information, potentially undermining public trust in the clearinghouse.

We study this problem using China's high school admissions as our leading application, an unusually rich setting for our analysis. Unlike elementary and middle school admissions, which are largely residency-based, high school admissions in China are test-based and selective: each year, within each city, middle school graduates participate in a centralized test, and access to prestigious high schools is highly competitive. Middle schools differ in quality, and students from lower-performing middle schools are at a disadvantage in this competition. To address this concern, the central government mandates local policymakers to exogenously allocate a substantial share, typically 50\% or more, of high school seats to graduating middle schools in proportion to their enrollments, creating a reserve system in which each student's type corresponds to her middle school.\footnote{See the linked official documents of China's central government in \citeyear{MOE2002Quota} and \citeyear{StateCouncil2014ExamReform}, which proposed and institutionalized this policy.} Admission outcomes are reported together with seat types (reserved or open), and policymakers disclose cutoff scores for both seat categories, exactly the disclosed information on which our verifiability concept rests.

A striking feature of this institutional landscape is that Chinese cities operate a wide variety of mechanisms, differing in whether reserved and open seats are allocated sequentially or simultaneously and in the order in which the two seat types are processed, yet policymakers rarely provide formal justification for these choices. To document this variation, we compile the admission mechanisms used by 35 major Chinese cities (\autoref{tab:mechanisms-in-china}). This diversity is the empirical puzzle that motivates our analysis: we show that verifiability provides a unifying lens that organizes observed designs in practice and yields concrete guidance for policymakers who must choose among them. The richness of this institutional context makes it an ideal setting for examining verifiability's implications for clearinghouse design.

We motivate our concept through the class of sequential mechanisms, which are prevalent in Chinese cities despite their known strategic drawbacks. Sequential mechanisms allocate reserved and open seats in two separate stages, each employing the student-proposing deferred acceptance (DA) algorithm; only students not admitted in stage 1 participate in stage 2. There are two variants, distinguished by the order in which seat types are allocated: one allocates reserved seats in stage 1 and open seats in stage 2, while the other reverses this order. Because each stage produces a stable outcome independently, students can verify their assigned schools within each stage by comparing their scores against the stage-specific cutoffs disclosed by policymakers. Crucially, the separation of the two stages automatically resolves the seat-type assignment: once a student confirms admission in a given stage, she knows the corresponding seat type without any additional information. Section~\ref{section:verifiability:concept} formalizes this cutoff-based reasoning and shows that these properties together imply a natural verifiability property of sequential mechanisms (Proposition~\ref{prop:verifiability:sequential}). However, sequential mechanisms have well-documented strategic flaws \citep{westkamp2013analysis,dur2019sequential,andersson2024sequential}: students may face a trade-off between securing a safer assignment in stage 1 and pursuing a more preferred one in stage 2, which can force nontrivial strategic behavior. This motivates the search for simultaneous mechanisms that preserve verifiability while providing stronger incentive properties.

The challenge for simultaneous mechanisms is that the joint allocation of reserved and open seats creates an ambiguity in seat-type assignment that sequential mechanisms avoid by design. When a student's score meets both the open-seat cutoff and her type's reserved-seat cutoff at her assigned school, her seat type cannot be inferred from the cutoffs alone. We address this by introducing two verification protocols for students under simultaneous mechanisms, each implementable using only a student's own score and the publicly disclosed cutoffs and analogous to one of the two stage orderings of the sequential mechanisms.
This restriction reflects the practical limits of both the information students can reasonably acquire and the computation they can reasonably perform, and is grounded in the verification procedures under sequential mechanisms already familiar in practice. Under the first \textit{reserve-first order}, a student infers that she is assigned to a reserved seat whenever her score meets the cutoff for her type's reserved seats; otherwise, she is assigned to an open seat. Under the second \textit{open-first order}, a student infers that she is assigned to an open seat whenever her score meets the open-seat cutoff; otherwise, she is assigned to a reserved seat. A mechanism is then \emph{verifiable} if every student can verify her own assignment consistently using one of these two orders; any outcome inconsistent with both orders is deemed unverifiable.

Verifiability is best understood as a governance property of the admission clearinghouse, with two complementary implications. First, it provides individual-level accountability: whenever the mechanism deviates from the announced rules, at least one student whose assignment is inconsistent will detect it. Any discrepancy can be expressed as a violation of the public cutoff-based rule, rather than as an opaque dispute over the internal allocation procedure. Second, verifiability is robust to heterogeneous information environments. In China, cities differ substantially in their disclosure practices: some disclose only cutoffs, others publish full profiles of students admitted through reserve quotas, and students sometimes privately collect information about peers. A verifiable mechanism provides consistent guarantees across all these environments: each student can verify her own assignment from the publicly disclosed cutoffs alone, and any student who acquires additional information about peers can apply the same verification rule to those peers' assignments and reach the same conclusion. In this sense, verifiability ensures individual verification under minimal disclosure and broader consistency under richer disclosure.

Section~\ref{section:simultaneous} presents the main results. We first show that verifiability essentially characterizes two school choice rules, $\C^{SimRO}$ and $\C^{SimOR}$ (Theorem~\ref{theorem:choicerule:characterization}). We then show that the DA mechanisms induced by these two rules are essentially the only simultaneous mechanisms satisfying individual rationality, strategy-proofness, and verifiability (Theorem~\ref{theorem:mechanism:characterization}). The first rule, $\C^{SimRO}$, assigns students to reserved seats before open seats, and is the unique verifiable rule consistent with the reserve-first order. It is not new: in the two-type setting with a single targeted group, it reduces to the choice rule of \citet{hafalir2013effective}, and in more general environments with multiple types, it coincides with the reserves rule of \citet{echenique2015how}. The second rule, $\C^{SimOR}$, is novel. It assigns seats as open seats first and uses reserved seats only when necessary, implemented through a within-school DA sub-procedure that treats applicants as preferring open seats to reserved seats. In the two-type setting, this construction is equivalent to filling open seats before reserved seats as studied by \citet{dur2018reserve}, who show that this rule maximizes the number of targeted students admitted within a broad class of choice rules. When there are multiple student types, however, $\C^{SimOR}$ is not equivalent to any known rule in the literature.

The characterization of verifiable rules is nontrivial, and the difficulty lies in the multi-type case. In the two-type setting, verifiable rules correspond exactly to the two natural precedence orders of seat types. With multiple student types, however, every choice rule based on a fixed precedence order of seats is unverifiable (Section~\ref{section:slot-specific}). This failure encompasses the broad class of slot-specific priority rules introduced by \citet{kominers2016matching} and its generalization with capacity transfers by \citet{avataneo2021slotspecific}, both of which contain all fixed precedence orderings as special cases and have been applied to a wide range of allocation problems. The novel content of the characterization is therefore $\C^{SimOR}$: its construction requires a DA sub-procedure that does not reduce to any fixed ordering of seat types, and establishing that it is, up to outcome equivalence, the unique verifiable rule consistent with the open-first verification order requires analysis specific to the multi-type case. Among outcome-equivalent rules, $\C^{SimOR}$ always minimizes the use of reserved seats; it assigns seats as reserved seats only when necessary.

The mechanism characterization builds on the approach of \citet{imamuraRecipe}, who show that if a choice rule axiom is \emph{punctual} and satisfies path independence and size monotonicity, then the corresponding DA mechanism is the unique mechanism satisfying individual rationality, strategy-proofness, and the mechanism-level extension of that axiom.\footnote{A school choice rule axiom is \emph{punctual} if it can be checked in each choice problem in isolation without reference to the other choice problems.} We extend their framework to accommodate seat types and verify that the required conditions hold in the extended setting; in particular, verifiability, formulated as a property of school choice rules, is punctual.

Among the two verifiable choice rules, we identify $\C^{SimOR}$ as particularly appealing: it minimizes the use of reserved seats and, in the two-type setting, maximizes the number of students who benefit from affirmative action. We view reserve quotas as a budget that permits schools to admit targeted students ahead of higher-ranked peers: $\C^{SimOR}$ spends this budget only when necessary, assigning reserved seats exclusively to students who cannot secure an open seat on merit, so that every reserved-seat assignment represents genuine affirmative action. Section~\ref{section:applicationtoChina} applies these results to interpret mechanism choices across Chinese cities and to discuss implications for other countries. Section~\ref{section:other:choicerule} examines the relationship between $\C^{SimOR}$ and related choice rules in the literature, including the de-reserve approach of \citet{aygun2023how} and the framework of \citet{abdulkadiroglu2025market}. Section~\ref{section:conclusion} concludes. The appendix contains proofs and additional results.

\paragraph{Related Literature}

Our paper contributes to the growing agenda on transparency, trust, and governance in the design and operation of marketplaces and allocation systems. Within this agenda, we study how the policymaker in a centralized admission clearinghouse can make affirmative action verifiable to its participants from minimal disclosed information. We organize the related work into three strands: the design of affirmative action through reserve systems, concepts of verification and transparency in mechanism design, and the role of processing orders in shaping admission outcomes.

Since the seminal work of \citet{abdulkadiroglu2003school}, a rich literature has studied the design of affirmative action through reserve systems \citep{kojima2012school,hafalir2013effective,ehlers2014school}; see \citet{sonmez2022affirmative}, \citet{aygun2023how}, and \citet{abdulkadiroglu2025market} for comprehensive surveys. We contribute to this literature by introducing verifiability as a mechanism design criterion. Our concept requires that students be able to verify both their assigned schools and their seat types using only their private scores and the publicly disclosed cutoffs. Seat-type verification is the key new element: without it, our concept reduces to a stability condition, which is too permissive to yield a meaningful characterization (Example~\ref{example:stability}).

The closest concept in the literature is the ``verifiability problem'' of \citet{hakimov2024improving}. Their paper takes the mechanism as given and asks which communication protocols allow students to verify their school assignments; verifiability is a property of the protocol. Our paper inverts the question: we fix the protocol (that is, policymakers disclose only seat-specific cutoffs) and ask which mechanisms are verifiable; verifiability is a property of mechanisms. Moreover, students must verify both assigned schools and seat types.

Other related concepts, including credibility \citep{akbarpour2020credible}, auditability \citep{grigoryan2024theory}, and transparency \citep{hakimov2024improving,moller2024transparent}, study whether policymakers' deviations from announced mechanisms can be detected from  participants' observations, and therefore treat verification as a commitment or monitoring device at the mechanism level. Our concept differs in two respects. First, we do not ask whether policymakers manipulate disclosed cutoffs; we take disclosed cutoffs as truthful and ask whether assignments are consistent with them. Second, we assume specific verification protocols; any outcome inconsistent with them is deemed unverifiable, even if a more elaborate investigation with more information might reveal consistency. This restriction reflects the practical limits of both the information students can reasonably acquire and the computations they can reasonably perform.

The auditability condition of \citet{pycia2024ordinal} also differs. It is used to characterize when certain desirable properties of a social choice or allocation rule can be certified from limited ordinal information. Our concern is not whether an outcome satisfies a welfare or efficiency benchmark, but whether an individual participant can verify her own school and seat-type assignment under a reserve system.

\citet{bonet2024explainable} introduce explainability for priority-based problems, focusing on justifying priority violations relative to feasible allocations; our concept concerns direct verification of assignments from disclosed information.

Our results are related to the reserves literature, but the connection is more nuanced than it may first appear. $\C^{SimRO}$ is equivalent to the reserve rule of \citet{echenique2015how} in the multi-type setting and reduces to the rule of \citet{hafalir2013effective} in the two-type setting. However, seat-type verification under $\C^{SimRO}$ does not follow immediately from this equivalence and requires additional analysis. $\C^{SimOR}$ is more fundamentally novel: in the two-type setting it coincides with the open-first processing order of \citet{dur2018reserve}, who show it maximizes the number of targeted students within a broad class of rules, but in the multi-type setting it cannot be represented as a fixed seat priority ordering. We prove that all choice rules based on fixed precedence orderings of seats, including the slot-specific priority rules of \citet{kominers2016matching} and the more general capacity-transfer rules of \citet{avataneo2021slotspecific}, fail verifiability when there are more than two student types. The de-reserve approach of \citet{aygun2023how}, applied to one unverifiable rule we study, yields a rule outcome-equivalent to $\C^{SimOR}$ but differs in construction.

The role of processing orders in shaping admission outcomes has been further analyzed by \citet{dur2020explicit}, \cite{pathak2024fair}, and \citet{pathak2023reversing,pathak2025immigration}. Most of the literature, including \citet{abdulkadiroglu2003school}, \citet{hafalir2013effective}, and \citet{ehlers2014school}, does not introduce seat types in outcomes, since students are indifferent over seats within a school; \citet{aygun2020dynamic,aygun2023affirmative} introduce seat types because participants have preferences over them. We introduce seat types to accommodate policy requirements even when students are indifferent, and show that this structure facilitates the analysis of affirmative action more broadly.

In concurrent work, \citet{durchinese} and \citet{dur2024who} study the score floor constraints imposed on reserved-seat students in Chinese cities.\footnote{Some Chinese cities require that the scores of students admitted to reserved seats cannot fall below the lowest score of those admitted to open seats at the same school by more than a fixed margin.
We exclude this detail from our analysis, since it is a secondary component of the system compared to those we study, and there is a trend in China towards eliminating this constraint. For example, Wuhan, a city discussed by both \cite{durchinese} and \cite{dur2024who}, has eliminated the floor constraint for reserved-seat students since 2024. See the \href{https://jyj.wuhan.gov.cn/zwdt/tsgg/202404/t20240415_2388662.shtml}{2024} document of Wuhan Municipal Education Bureau.} \citet{durchinese} classify mechanisms by whether the constraint is fixed or dynamic and show that fixed constraints yield substitutable choice rules while dynamic constraints do not. \citet{dur2024who} analyze the optimal allocation of score bonuses to eligible students and document the opacity and corruption vulnerability of China's early mechanisms, which allowed middle schools to allocate reserved seats through pre-test procedures. The subsequent reform toward the centralized, test-based mechanisms we study underscores the importance of transparency in the implementation of affirmative action.

\section{The Model} \label{section:model}

We define the controlled school choice model \citep{ehlers2014school,echenique2015how} extended to incorporate seat types in outcomes. The model consists of the following elements:
\begin{itemize}
	\item A finite set of students $\I=\{i_1,\cdots,i_N\}$, with a typical student denoted by $i$ or $j$;
	
	\item A finite set of schools $\S=\{s_1,\cdots,s_K\}$, with a typical school denoted by $s$ or $s'$;

	\item A profile of school capacities $q=\{q_s\}_{s\in \S}$, where $q_s \in \mathbb{N}\cup \{0\}$ denotes the number of seats at school $s$. Let $\emptyset$ denote a null school with an unlimited capacity; 
	
	\item A profile of student preferences $P_{\I}=\{P_i\}_{i\in \I}$, where $P_i$ denotes the strict preference relation of student $i\in \I$ over $ \S\cup \{\emptyset\} $. We write $s R_i s^{\prime} $ if $ s P_is^{\prime}$ or $s = s^{\prime}$. A school $s$ is \textit{acceptable} to student $i$ if $ s P_i \emptyset $, and otherwise we call it \textit{unacceptable} to $ i $;
	
	\item  A profile of school priorities $\succ_{\S}=\{\succ_s\}_{s\in \S}$, where $\succ_s$ is the strict priority order used by school $s\in \S$ to rank students. Each student $i$ has a score $\pi_{i,s} \in \mathbb{N}$ at each school $s$ such that, for any two students $i$ and $j$, $i \succ_s j$ if and only if $ \pi_{i,s}>\pi_{j,s}$; set $\pi_{i,\emptyset}=0$ for all $i\in \I$;
	
	\item A type space for students $\M=\{m_1,m_2,\ldots,m_L \}$ and a function $\tau: \I \rightarrow \M$ such that, for each student $i$, $\tau(i)$ is $i$'s type. For each $m\in \M$, let $\I^m= \tau^{-1}(m)$ denote the set of type-$m$ students. For each $ I\in 2^\I $, let $ I^m= I\cap \I^m $;

	\item A profile of \textit{reserve quotas} $\{q_s^r\}_{s\in \S}$, where $ q_s^r=\{q_s^m\}_{m\in \M}$, such that $ q^m_s\in \mathbb{N}\cup \{0\} $ and $\sum_{m\in \M} q_s^m < q_s$. For each school $s$, $ q^m_s$ is the quota of \textit{reserved seats} for type-$ m $ students, and $ q^o_s= q_s-\sum_{m\in \M} q_s^m $ is the quota of \textit{open seats} for all students. We assume that $q^m_s < |\I^m|$ for every school $ s $ and every student type $ m $.\footnote{This assumption ensures that reserved seats are binding; if all students of a given type fit within the reserve quota, affirmative action is non-binding and the distinction between seat types becomes trivial.} 
\end{itemize}

In the context of China's high school admissions, $\S$ represents the set of high schools, $\M$ represents the set of middle schools from which students graduate, $q^m_s$ represents the number of reserved seats at high school $s$ for middle school $m$, and $ \pi_{i,s} $ represents student $ i $'s score in admission tests. Though Chinese students are typically ranked according to their test scores,\footnote{In practice, score ties are broken according to exogenous rules, so students are strictly ranked.} our model allows schools to rank students using different priority orders. 


\paragraph{Outcome} In the literature, a matching is often defined as a function that assigns students to schools without exceeding the capacity of each school. However, such a definition is insufficient to describe the admission outcome for our purpose. In our applications, policymakers are required to reveal the seat types admitted students occupy. Students occupying reserved seats are viewed as beneficiaries of affirmative action, who are protected from justified envy from students of other types. Different allocations of reserved seats indicate different implementations of affirmative action parameters. 

Formally, an admission outcome in our model is defined as follows.

\begin{definition}\label{definition:matching}
	An admission \textbf{outcome} $ \mathbf{\mu} $ is a subset of $ (\S\cup\{\emptyset\})\times \I \times \mathcal{T} $, where $ \mathcal{T}=\{r(eserved),o(pen)\} $ denotes the set of two seat types, such that
	
	\begin{enumerate}
		\item[(1)] $ |\{(s,i,t)\in \mu:s\in \S\cup \{\emptyset\},t\in \mathcal{T}\}|= 1 $ for all $ i\in \I $,
		
		\item[(2)]  $ |\{(s,i,t)\in \mu:i\in \I, t\in \mathcal{T}\}|\le q_s $ for all $ s\in \S $,
		
		\item[(3)] $ |\{(s,i,r)\in \mu:i\in \I^m\}|\le q^m_s $ for all $ s\in \S $ and all $ m\in \M $.
	\end{enumerate}
\end{definition}
In Definition \ref{definition:matching}, for any student $ i $, if $ (s,i,t)\in \mu $ for some $ s\in \S $ and $ t\in \mathcal{T} $, it means that $ i $ is admitted to a type-$t$ seat at school $ s $. If $ (\emptyset,i,t)\in \mu $ for some $ t\in \mathcal{T} $, it means that $ i $ is unmatched, in which case the type of her seat is irrelevant; for convenience, we say that $i$ occupies an open seat at $ \emptyset $.  Condition (1) says that each student is admitted to exactly one school, possibly the null school.  Condition (2) says that each school's capacity is respected. Condition (3) says that, for each $m\in \M$, each school $ s $ provides at most $ q^m_s $ reserved seats to type-$m$ students. 

Given an outcome $ \mu $, for each school $s$, let $\mu(s)= \{(s,i,t)\in \mu:i\in \I,t\in \mathcal{T}\}$, and for each student $i$, let $\mu(i)= \{(s,i,t)\in \mu:s\in \S\cup\{\emptyset\},t\in \mathcal{T}\}$. When there is no confusion, we also use $ \mu(s) $ to denote the set of students admitted to school $ s $ and use $ \mu(i) $ to denote the school admitting student $ i $. Given a school $ s $, for each seat type $ t\in \T $, let $ \mu^t(s) $ denote the set of admitted students who occupy type-$ t $ seats. Similarly, for each $ m\in \M $, let $ \mu^m(s) $ denote the set of type-$m$ students admitted to $ s $. Additionally, we define $ \mu^{rm}(s)=\mu^r(s)\cap \mu^m(s) $, which is the set of type-$m$ students who occupy reserved seats at $ s $. For each $ i\in \I $, let $ t^{\mu}_{i} $ denote the type of $ i $'s seat in $ \mu $.

Given an outcome $ \mu $, by ignoring students' seat types, we obtain a matching in the standard definition. Formally, a (standard) \textbf{matching} is a function $ u: \I\cup \S \rightarrow 2^{\I} \cup \S \cup \{\emptyset\} $ such that, for each $ i \in \I$, $ u(i)\in \S\cup \{\emptyset\} $, for each $ s\in \S $, $ u(s)\in 2^{\I} $ with $ |u(s)|\le q_s $, and for each student-school pair $ (i,s) \in \I\times \S $, $ u(i)=s $ if and only if $ i\in u(s) $. 
We say that a  matching $ u $ is induced by an outcome $ \mu $ if, for each $ s\in \S $, $ u(s)= \{i\in \I: (s,i,t)\in \mu \text{ for some } t\in \mathcal{T}\} $. A matching can be induced by multiple outcomes that differ in the allocation of seat types.


\paragraph{Stability} In the usual definition of stability proposed in the literature (e.g., \citealp{hafalir2013effective}), a school's reserve quota $q^m_s$ is interpreted as a ``floor'' for affirmative action in the sense that, as long as there are enough applicants from $ \I^m $, the school must admit at least as many as the reserve quota, even if this creates priority violations. On the other hand, $q^m_s$ is also treated as a ``ceiling'' for affirmative action in the sense that, once the school admits more type-$m$ students than the reserve quota allows, the admitted  students are no longer allowed to initiate priority violations. However, this treatment of reserve quotas does not align with the practices in many countries. In China, it is commonly observed that a high school admits more students from a middle school than the reserve quota allows, yet the high school still uses the reserve quota to prioritize students from that middle school.

In this paper, we interpret a reserve quota $q^m_s$ as a ``budget'' for priority violations that school $s$ can create to favor students from $ \I^m $; students from $ \I^m $ who benefit from such violations occupy reserved seats. By specifying seat types, our approach can accommodate various usages of the ``budget''.

Specifically, in an outcome $ \mu $, a student $ i $ is said to have \textbf{justified envy} towards another student $ j $  if $i$ prefers $j$'s assigned school to her own and has higher priority than $j$ for that school, and either they have the same student type, or $ j $ occupies an open seat; that is, $\mu(j) P_i \mu(i)$, $i \succ_{\mu(j)}j$, and [$ \tau(i)= \tau(j) $ or $ t^{\mu}_j=o $].


An outcome $ \mu $ is \textbf{individually rational} if no student is assigned to an unacceptable school; that is, for all $ i\in \I $, $ \mu(i) R_i \emptyset $.

An outcome $ \mu $ is \textbf{non-wasteful} if every school preferred by a student over her assignment allocates all of its seats to students; that is, for all $ i\in \I $ and all $ s P_i \mu(i) $, $ |\mu(s)|=q_s $.

Finally, an outcome $\mu$ is \textbf{reserve-non-wasteful} if every school preferred by a student over her assignment has used up its reserve quota for that student's type; that is, for all $ i\in \I $ and all $ s P_i \mu(i) $, $ |\mu^{r\tau(i)}(s)|=q^{\tau(i)}_s $. Schools are viewed as obligated to allocate reserved seats, up to the reserve quota, whenever eligible students request such seats.

\begin{definition}
	An outcome $ \mu $ is \textbf{stable} if it is individually rational, non-wasteful, reserve-non-wasteful, and does not create justified envy.
\end{definition}

Example \ref{example:stability} illustrates the flexibility of our definition in accommodating different allocations of reserved seats.

\begin{example}\label{example:stability}
	Consider a school $s$ and six students $\I = \{i_1,i_2,i_3,i_4,i_5,i_6\}$ who all accept $s$. Students belong to two types, $\I^{m_1}=\{i_1,i_2,i_3\}$ and $\I^{m_2}=\{i_4,i_5,i_6\}$. Suppose $q_s=4$ and $q_s^{m_1}=q_s^{m_2}=1$. The priority order used by $ s $ is $i_1 \succ i_4 \succ i_2 \succ i_3 \succ i_5 \succ i_6$. 
	
	Since $q_s^{m_1}=q_s^{m_2}=1$, in any stable outcome, $s$ must admit at least one student from each type. By exploring all possible assignments of the two types of seats, we obtain five stable outcomes, which are listed in the following table. In the table, $s^o$ denotes an open seat;  $s^{m_1}$ and $s^{m_2}$ denote two reserved seats, one for each type.  
	
	\begin{table}[!h]
		\centering
		\begin{tabular}{c|cccc}
			& $s^o$ & $s^o$ & $s^{m_1}$ & $s^{m_2}$ \\ \hline
			$\mu_1$ & $i_2$  & $i_3$ & $i_1$ & $i_4$ \\
			$\mu_2$ & $i_1$  & $i_2$ & $i_3$ & $i_4$ \\
			$\mu_3$ & $i_1$  & $i_3$ & $i_2$ & $i_4$ \\
			$\mu_4$ & $i_1$  & $i_4$ & $i_2$ & $i_5$ \\
			$\mu_5$ & $i_4$  & $i_2$ & $i_1$ & $i_5$
		\end{tabular}
	\end{table}
	
	Observe that $\mu_1(s) = \mu_2(s) = \mu_3(s) =  \{i_1, i_2, i_3, i_4\}$ and $\mu_4(s) = \mu_5(s) = \{i_1, i_2, i_4, i_5\}$. Since $\mu_1$, $\mu_2$, and $\mu_3$ admit the top four students in the priority order, no students benefit from affirmative action, although reserved seats are assigned. In contrast, $\mu_4$ and $\mu_5$ admit the top three students and $ i_5 $. So, $ i_5 $ benefits from affirmative action.\footnote{In $\mu_4$, $i_2$ occupies a reserved seat; in $\mu_5$, $i_1$ occupies a reserved seat, to fulfill the reserve-non-wastefulness requirement. But they do not benefit from affirmative action.} In the conventional definition, $\mu_4$ and $\mu_5$ are not viewed as stable because, given that $i_4$ is admitted and meets the reserve quota requirement for her type, $i_5$ is not protected from priority violations. But these two outcomes are stable in our definition.
\end{example}


\paragraph{Mechanism} When defining mechanisms, we assume that only students' preferences are unknown to policymakers. A \textbf{mechanism} specifies a strategy space $\Sigma_i$ for each student $i$ along with a function $\psi$ that selects an outcome for each strategy vector $(\sigma_1,\sigma_2,\ldots,\sigma_n)\in \times_{i\in \I}\Sigma_i$. A mechanism is said to satisfy an axiom defined for an outcome if the outcome it finds for every strategy vector satisfies the same axiom. 
When $\Sigma_i$ equals the domain of preferences for all $i\in \I$, we call $\psi$ a \textbf{direct mechanism}. In the paper, we assume that all strict preferences are possible and use $\mathcal{R}$ to denote this domain.  A direct mechanism $\psi$ is \textbf{strategy-proof} if, for any preference profile $P_\I\in \mathcal{R}^\I$, any student $i\in \I$, and any possible manipulation $P'_i\neq P_i$, $\psi_i(P_\I) R_i \psi_i(P'_i,P_{-i})$. Two direct mechanisms are called \textbf{equivalent} if they find an identical outcome for every preference profile.

In the paper, we will discuss mechanisms that use different strategy spaces. Specifically, we will study mechanisms that require students to submit a \textit{pair} of preference lists and that admit students to schools in two stages. We refer to them as \textbf{sequential mechanisms}. We will also study direct mechanisms and refer to them as \textbf{simultaneous mechanisms}, since they require students to submit a single preference list and assign them in one stage.

\section{Verifiability of School and Seat Assignments}\label{section:verifiability:concept}

If policymakers were able to disclose information without constraint, then students could verify their assignments under any mechanism. In the extreme, if all details of the admission process were disclosed, students could replicate the entire procedure on their own computers and verify the assignments of all students. Such a level of transparency is clearly unrealistic. In practice, policymakers face institutional and cultural constraints that limit what can be disclosed. Motivated by the practices in several countries, we assume that policymakers publicly disclose no information beyond school cutoffs. Students must rely only on their private scores and the publicly disclosed cutoffs to verify their assignments.

Given an outcome $\mu$, a \textbf{school cutoff profile} $c(\mu)=\bigl(\{c^m_s\}_{m\in\mathcal{M}},c^o_s\bigr)_{s\in\mathcal{S}\cup\{\emptyset\}}$ specifies, for each school $s$, the cutoff for type-$m$ reserved seats, $c^m_s$, and the cutoff for open seats, $c^o_s$. We focus on non-wasteful outcomes: whenever a school has unfilled seats, they are available to any student, so the open-seat cutoff is set to zero. This yields:
\begin{equation}\label{equation:cutoff:new}
\scalebox{.9}{$c^m_s=\begin{cases}
        \min_{i\in \mu^{rm}(s)} \pi_{i,s}, & \text{if }|\mu^{rm}(s)|=q^m_s>0; \\
        0, & \text{if } q^m_s>0 \text{ and }|\mu^{rm}(s)|<q^m_s;\\
        +\infty, & \text{if } q^m_s=0.
    \end{cases}
    \quad
    c^o_s=\begin{cases}
        \min_{i\in \mu^{o}(s)} \pi_{i,s}, & \text{if }|\mu(s)|=q_s; \\
        0, & \text{otherwise}.
    \end{cases}
    $}
\end{equation}
The cutoffs for all seats at the null school are set to zero.

Given a cutoff profile $c(\mu)$, the set of \textbf{achievable schools} for student $i$ in outcome $\mu$ is
$$\mathcal{S}^\mu_i = \Bigl\{s\in\mathcal{S}\cup\{\emptyset\}\colon \pi_{i,s}\ge \min\bigl\{c^{\tau(i)}_s,\, c^o_s\bigr\}\Bigr\}.$$
School $s$ is achievable to student $i$ if her score meets at least one of the cutoffs at $s$. We use $ \max (P_i, \S^\mu_i) $ to denote the best school among $ \S^\mu_i $ according to $ P_i$. 

In the classical school choice framework, \cite{azevedo2016supply} show that every stable matching corresponds to a vector of market clearing school cutoffs, and vice versa. This remains true in school choice with a reserve system.

\begin{lemma}\label{lemma:stable}
    Given students' submitted preferences $P_\mathcal{I}$ and an outcome $\mu$,
    $\mu(i)=\max(P_i,\mathcal{S}^\mu_i)$ for every student $i$
    if and only if $\mu$ is stable.
\end{lemma}

Lemma~\ref{lemma:stable} shows that every student can verify her school assignment using only her own score and the disclosed cutoffs. So, the true challenge is the verification of \textit{seat-type} assignment: when a student's score meets both $c^{\tau(i)}_s$ and $c^o_s$ at her assigned school $s$, the cutoffs alone do not determine whether she occupies a reserved or an open seat.

Section~\ref{section:sequential} shows that sequential mechanisms resolve this ambiguity naturally through their two-stage structure, and characterizes the resulting verification procedures. Although sequential mechanisms are strategically complex, their verification procedures serve as templates for the concept of verifiability for simultaneous mechanisms introduced in Section~\ref{section:verifiability:simultaneous:new}.

\subsection{Verifiability in Sequential Mechanisms}\label{section:sequential}

Sequential mechanisms allocate reserved seats and open seats across all schools in two distinct stages. Each student $i$ is required to report a pair of preference lists $ (P^r_i, P^o_i) $ before the mechanisms are run: $ P^r_i $ is for the reserved-seat stage and $ P^o_i $ is for the open-seat stage. Depending on the order of the two stages, we call the two variants \textbf{sequential-reserve-open} (SeqRO) and \textbf{sequential-open-reserve} (SeqOR) respectively. SeqRO allocates reserved seats in stage 1 and open seats in stage 2, and SeqOR reverses the order.  In both mechanisms, each stage runs DA using participating students' reported preferences and schools' quotas. 

\begin{itemize}
	\item \textbf{SeqRO}: 	Stage 1 runs DA using students' reported $ P^r_i $ and schools' reserve quotas $q^m_s$. In each step, each unassigned student $i$ applies to the highest-ranked acceptable school in $ P^r_i $ that has not rejected her; each school $s$ tentatively admits the highest-priority applicants of each type $ m $ from new applicants in this step and tentatively admitted applicants from the previous step, up to quota $q^m_s$. Stage 1 stops when each student either has been admitted to an acceptable school or has been rejected by every acceptable school. At each school $s$, admitted students occupy reserved seats, and any unfilled reserve quota is added to its open-seat quota for stage 2. Unassigned students in stage 1 participate in stage 2. 	
	
	Stage 2 runs DA using participating students' reported $ P^o_i $ and schools' open-seat quotas, similar to Stage 1 except that, in each step, each school $s$ tentatively admits the highest-priority applicants regardless of their type up to the open-seat quota. Admitted students occupy open seats. Students not admitted in both stages are unmatched.

	\item \textbf{SeqOR}: The order of the two stages of SeqRO is reversed, with the difference that stage 1 runs DA using schools' original open-seat quotas $q^o_s$.
\end{itemize}

Under SeqRO, the unassigned reserved seats in stage 1 are allocated as open seats in stage 2. However, in SeqOR, the reserve quotas are set aside when open seats are allocated in stage 1. Consequently, the unassigned reserved seats in stage 2 of SeqOR can be wasted. 

Under both mechanisms, since each stage runs DA, the outcome in each stage is stable with respect to participating students' submitted preferences and is therefore characterized by school cutoffs. Students can thus verify their school assignments using only their private scores and publicly disclosed school cutoffs in both stages. A student first verifies whether she is eligible for any school in stage 1, and if so, whether she is assigned to the best such school according to her submitted preferences. If she is not eligible for any school in stage 1, she then verifies whether she is eligible for any school in stage 2, and if so, whether she is assigned to the best such school according to her submitted preferences. Once a student's assigned school is verified in either stage, her seat-type assignment is automatically verified.

The above observation is simple, but we formalize it for later reference. Under SeqRO, the school cutoffs are defined by \autoref{equation:cutoff:new}: unfilled reserved seats are converted to open seats, so the cutoff convention follows the non-wasteful formula. Under SeqOR, the reserved-seat cutoff is calculated as in \autoref{equation:cutoff:new}, but because reserved seats are set aside when open seats are allocated, the open-seat cutoff depends only on the exogenous open quota: $c^o_s=\min_{i\in \mu^{o}(s)} \pi_{i,s}$ if $|\mu^o(s)|=q^o_s$, and $c^o_s=0$ otherwise.

Given students' submitted preferences $ \{P^r_i, P^o_i\}_{i\in \I} $ and the school cutoff profile $ c(\mu)$ in an outcome $ \mu $, we define $ \S^o_i= \{s\in \S: \pi_{i,s}\ge c^o_s,~sP^o_i \emptyset \} $ and $ \S^r_i=\{s\in \S: \pi_{i,s}\ge c^{\tau(i)}_s,~sP^r_i \emptyset\} $ as the sets of schools achievable to a student $ i $ in the two stages. 

The following result formalizes students' verification order under the two mechanisms.

\begin{proposition}\label{prop:verifiability:sequential}
	Under either of the two sequential mechanisms, let $ \{P^r_i, P^o_i\}_{i\in \I} $ be students' submitted preferences and $\mu$ the matching outcome. Then, for every student $ i $:
	\begin{table}[!h]
			\begin{subtable}[b]{.4\linewidth}
				\begin{flushleft}
					\scalebox{.85}{$(\mu(i),t^\mu_i)=\begin{cases}
							(\max (P^r_i, \S^r_i),r), & \text{if }\S^r_i\neq \emptyset; \\
							(\max (P^o_i, \S^o_i),o), & \text{if } \S^r_i= \emptyset \text{ and }\S^o_i\neq \emptyset;\\
							(\emptyset,o), & \text{if } \S^r_i= \emptyset \text{ and }\S^o_i=\emptyset.
						\end{cases}
						$}
				\end{flushleft}
				\subcaption{Under SeqRO}
			\end{subtable}
			\begin{subtable}[b]{.63\linewidth}
				\begin{flushright}
					\scalebox{.85}{$(\mu(i),t^\mu_i)=\begin{cases}
							(\max (P^o_i, \S^o_i),o), & \text{if }\S^o_i\neq \emptyset; \\
							(\max (P^r_i, \S^r_i),r), & \text{if } \S^o_i= \emptyset \text{ and }\S^r_i\neq \emptyset;\\
							(\emptyset,o), & \text{if } \S^o_i= \emptyset \text{ and }\S^r_i=\emptyset.
						\end{cases}
						$}
				\end{flushright}
				\subcaption{Under SeqOR}
			\end{subtable}
		\end{table}
\end{proposition}

We omit the proof of the above proposition since it resembles Lemma \ref{lemma:stable}.

Although sequential mechanisms have intuitive verifiability properties, they are manipulable and might be strategically complex for students. Students often confront a trade-off between securing a safe assignment in stage 1 and risking the possibility of obtaining a better assignment in stage 2. Appendix \ref{appendix:NE:sequential} presents a Nash equilibrium (NE) analysis of sequential mechanisms under the complete information assumption (Proposition \ref{proposition:NE of the sequential-reserve-open}). The set of NE outcomes of SeqRO coincides with the set of stable matchings; in contrast, under SeqOR, not every stable matching is attainable as a NE outcome, and wasteful NE outcomes may arise. However, incomplete information prevails in practice and complicates students' strategies. This motivates our search for strategy-proof simultaneous mechanisms in Section \ref{section:simultaneous}.

\subsection{Definition of Verifiability for Simultaneous Mechanisms}\label{section:verifiability:simultaneous:new}

Although sequential mechanisms are verifiable by design, they are strategically complex. Simultaneous mechanisms address this by assigning all seats in a single stage with a single preference list. However, unlike sequential mechanisms, they must resolve seat-type assignment without the benefit of separate stages: once a student's school assignment is verified, the cutoffs alone may not determine whether she occupies a reserved or an open seat. So we need an exogenous rule to resolve the ambiguity.

We propose a verification protocol with two variants that mirror the two verification orders under sequential mechanisms: the \textbf{reserve-first order} and the \textbf{open-first order}. Under the reserve-first order, which mirrors SeqRO, a student infers that she occupies a reserved seat if her score meets her type's reserved-seat cutoff at her assigned school, and an open seat otherwise. Under the open-first order, which mirrors SeqOR, she infers that she occupies an open seat if her score meets the open-seat cutoff, and a reserved seat otherwise. Both orders require only the student's own score and the two seat-type cutoffs at her assigned school. Importantly, they are the only orders that naturally extend the familiar verification protocols under sequential mechanisms. The paper asks which simultaneous mechanisms preserve those familiar verification protocols.

We formally define verifiability for an outcome as follows.

\begin{definition}
	Given students' submitted preferences $ P_\I $, an outcome $\mu$ is \textbf{verifiable} if, for every student $ i $, $
	\mu(i)= \max (P_i,\S^\mu_i)$, and $\mu$ follows one of the two verification orders:
	\begin{enumerate}
		\item \textbf{Reserve-first order}: for every matched $ i $, $ t^{\mu}_i= r $ if $ \pi_{i,\mu(i)}\ge c^{\tau(i)}_{\mu(i)} $ and otherwise $ t^{\mu}_i= o $.

		\item \textbf{Open-first order}: for every matched $ i $, $ t^{\mu}_i=
		o$ if $ \pi_{i,\mu(i)} \ge c^o_{\mu(i)} $ and otherwise $ t^{\mu}_i=
		r $.
	\end{enumerate}
In the first case, we call $ \mu $ \textit{reserve-first verifiable}; in the second case, \textit{open-first verifiable}.
\end{definition}

We then define verifiability for a simultaneous mechanism as follows.

\begin{definition}\label{definition:mechanism:verifiable}
	A simultaneous mechanism is \textbf{(reserve-first/open-first) verifiable} if the outcome it finds for every preference profile is (reserve-first/open-first) verifiable.
\end{definition}

The following example illustrates the concept by contrasting a verifiable outcome with an unverifiable one.

\begin{example}[A verifiable and an unverifiable seat-type assignment]\label{example:verifiability:contrast}
Consider a single school $s$ with capacity $q_s=3$, consisting of $q^o_s=2$ open seats and $q^m_s=1$ reserved seat for a single targeted type $m$. Three students $i_1,i_2,i_3\in\I^m$ find school $s$ acceptable, with scores $\pi_{i_1,s}=10>\pi_{i_2,s}=7>\pi_{i_3,s}=4$. All three are admitted to $s$, so the school assignment is the same in both outcomes below; the only question is which seat type each student occupies.

\textbf{A verifiable outcome}: Suppose $i_1$ and $i_2$ occupy open seats and $i_3$ occupies the reserved seat. By \autoref{equation:cutoff:new}, the cutoffs are $c^o_s=\min\{\pi_{i_1,s},\pi_{i_2,s}\}=7$ and $c^m_s=\pi_{i_3,s}=4$. This outcome is open-first verifiable: a student occupies an open seat exactly when her score meets $c^o_s=7$, which holds for $i_1$ and $i_2$ and fails for $i_3$. Each student can therefore reconstruct her seat type from her own score and the two cutoffs.

\textbf{An unverifiable outcome}: Suppose instead that $i_1$ and $i_3$ occupy the open seats and $i_2$ occupies the reserved seat. The cutoffs become $c^o_s=\min\{\pi_{i_1,s},\pi_{i_3,s}\}=4$ and $c^m_s=\pi_{i_2,s}=7$. The open-first order fails: it predicts that $i_2$, whose score $7$ meets $c^o_s=4$, occupies an open seat, contradicting her reserved seat. The reserve-first order also fails: it predicts that $i_1$ occupies a reserved seat, contradicting her open seat. Because the reserved-seat holder $i_2$ has a score strictly between those of the two open-seat holders, no cutoff-based rule can rationalize the labeling, and a student cannot tell from the cutoffs alone whether the reserved seat is correctly assigned.
\end{example}

\section{Verifiable and Strategy-proof Simultaneous Mechanisms} \label{section:simultaneous}

For simultaneous mechanisms, verifiability implies stability because the definition of verifiability incorporates a cutoff characterization of outcomes. Since we also seek strategy-proofness, this motivates our focus on the class of DA mechanisms, which differ in the choice rules used by schools. We therefore begin by defining choice rules.

A standard choice rule in the literature specifies the set of students chosen by a school from every potential set of applicants. Formally, a \textbf{standard choice rule} for a school $s$ is a function $C: 2^{\I}\rightarrow 2^{\I}$ such that, for each $ I\in 2^{\I}  $, $C(I)\subseteq I $ and $ |C(I)|\le q_s $. For each $ m\in \M $, let $C^m(I)$ denote the set of type-$ m $ chosen students.

In this paper, a \textbf{choice rule} $ \C $ must additionally specify the seat types assigned to chosen students. To simplify the definition, we treat school $ s $ and the set of applicants $ I $ as forming an artificial economy in which all applicants regard school $ s $ as acceptable. We then require  $\C(I)$ to be a well-defined outcome in this economy. This requirement ensures that $\C(I)$ respects the quotas for all seat types at school $ s $.
Let  $ \C\langle I \rangle $ denote the set of chosen students under $ \C(I) $.  For each $ t\in \T $, let $\C^t\langle I \rangle$ denote the set of chosen students who occupy type-$t$ seats. For each $ m\in \M $, let $\C^{tm}\langle I \rangle=\C^t\langle I \rangle\cap I^m$. For each chosen student $ i $, let $t^{\C(I)}_i$ denote her assigned seat type. Throughout the paper, we use the calligraphic font $ \C $ to denote a choice rule specifying seat types, and use $ C $ to denote a standard choice rule that does not specify seat types.
We say that $ C $ is induced by $ \C $ if, for all $ I\in 2^{\I}  $, $ C(I)= \C\langle I \rangle $.

We extend the concept of verifiability to choice rules.

\begin{definition}\label{definition:choicerule:verifiable}
	A choice rule $\C$ for a school $ s $ is \textbf{(reserve-first/open-first) verifiable} if, for every $I\in 2^{\I}$, $\C(I)$ is a (reserve-first/open-first) verifiable outcome in the artificial economy consisting of school $s$ and students $I$.
\end{definition}

It is well known that substitutability of choice rules is sufficient and almost necessary for the functioning of DA and for the stability of its outcome. The definition of substitutability does not refer to students' seat types: a standard choice rule $C$ is \textbf{substitutable} if, for all $I\in  2^\I$ and all distinct $i,j\in I$, $i \in C(I)$ implies $i \in C(I\backslash\{j\})$; $ C $ is \textbf{$ q $-acceptant} if, for all $I\in  2^\I$, $|C(I)| =\min\{q, |I|\}$. A choice rule $ \C $ is said to satisfy these properties if the induced $ C $ satisfies them.

The following result is the foundation for our approach.

\begin{lemma}\label{lemma:verifiability}
	If every school uses a choice rule that is substitutable and verifiable, then the corresponding DA mechanism is individually rational, strategy-proof, and verifiable.\footnote{When all schools use substitutable choice rules, DA runs as follows: in each round $ \ell $, each unassigned student applies to the highest-ranked school in her preferences that has not rejected her;  facing new applicants in this round and tentatively admitted students in the previous round, the total set of which is denoted by $ I^\ell_s $, each school $ s $ admits students in $ \C_s(I^\ell_s) $. The algorithm stops when all students are either admitted or have been rejected by all acceptable schools.}
\end{lemma}

Section \ref{section:choicerule:characterization} characterizes verifiable choice rules. Section \ref{section:mechanism:characterization} characterizes verifiable and strategy-proof simultaneous mechanisms. We show that these mechanisms must be DA associated with the choice rules characterized in Section \ref{section:choicerule:characterization}. So, our restricted attention to the class of DA mechanisms at the outset is without loss of generality.

\subsection{Characterization of Verifiable Choice Rules}\label{section:choicerule:characterization}

We first characterize verifiable choice rules. By Definition \ref{definition:choicerule:verifiable}, if a choice rule $\C$ is reserve-first verifiable, then, in any outcome $\C(I)$, students of each type who occupy reserved seats must have higher priority than students of the same type who occupy open seats. Moreover, the reserved seats for a given type must be exhausted before any student of that type is assigned to an open seat. If the reserved seats for some type are not exhausted, then any unassigned reserved seats must have been converted into open seats, provided that some students are rejected. This observation naturally leads to the following choice rule, denoted by $\C^{SimRO}$.
\begin{itemize}
	\item  For each $I\in 2^{\I}$, $\mathbf{\C^{SimRO}}$ selects students as follows:
	 In the first step, for each $ m\in \M $, allocate reserved seats to the highest-ranked applicants from $ I^m $ up to the reserve quota $ q^m_s $. In the second step, all remaining seats are treated as open seats and allocated to the highest-ranked students among remaining applicants, regardless of their type.
\end{itemize}
We call the corresponding DA mechanism \textbf{simultaneous-reserve-open} (SimRO). Theorem \ref{theorem:choicerule:characterization} proves that $\C^{SimRO}$ is the only reserve-first verifiable choice rule.

On the other hand, if a choice rule $\C$ is open-first verifiable, then, in any outcome $\C(I)$, students occupying open seats must have higher priority than those occupying reserved seats, regardless of their type. This suggests that such a choice rule should assign open seats before reserved seats. We now define a choice rule inspired by this idea, which can be viewed as a natural counterpart of $\C^{SimRO}$. The rule first assigns students to open seats up to the open quota, and then assigns the remaining students to reserved seats for their respective types up to the corresponding reserve quotas. A subtlety arises when the reserved seats for some type are not filled by targeted students: to ensure non-wastefulness, the unfilled reserved seats should be made available again as open seats. This leads to the choice rule denoted by $\C^{SimORO}$.
\begin{itemize}
	\item For each $I\in 2^{\I}$, $\mathbf{\C^{SimORO}}$ selects students as follows: In the first step, allocate open seats to the highest-ranked students, regardless of their type, up to the open quota $q^o_s$. In the second step, for each $ m\in \M $, allocate reserved seats to the highest-ranked remaining students from $ I^m $, up to the reserve quota $ q^m_s $. In the last step, if there exist unfilled reserved seats, they are treated as open seats and allocated to the highest-ranked remaining students, regardless of their type.
\end{itemize}
We call the corresponding DA mechanism  \textbf{simultaneous-open-reserve-open} (SimORO). As discussed in Section \ref{section:applicationtoChina}, both SimRO and SimORO are used in China's high school admissions.

However, unlike $\C^{SimRO}$, $\C^{SimORO}$ is not open-first verifiable. It is not surprising because in $\C^{SimORO}$, reserved seats might be assigned in a step between steps in which open seats are assigned. As a result, a student occupying a reserved seat may be ranked between two students occupying open seats. Section \ref{section:slot-specific} shows that all choice rules with fixed precedence orders of reserved and open seats are not verifiable. 

Now, we are ready to present our characterization of open-first verifiable choice rules. We define a specific choice rule, $\C^{SimOR}$, which selects students by running DA in an artificial economy consisting of school $ s $ and the set of applicants $ I $. Different from the artificial economy previously discussed, here, we regard ``open seats'' (denoted by $ s^o $) and ``reserved seats'' (denoted by $ s^r $) as two contract terms and assume that all students prefer $ s^o $ over $ s^r $.
\begin{itemize}
	\item  For each $I\in  2^{\I}$, $\mathbf{\C^{SimOR}}$ selects students as follows: In the artificial economy consisting of school $s$ and students $I$, run DA by letting all students apply to $ s^o $ before applying to $s^r$. In each step, after receiving applications, for each $ m\in \M $, $s$ first assigns reserved seats to the highest-ranked students from $I^m$ who apply to $s^r$, up to the quota $q_s^m$. Then, $s$ allocates the remaining seats as open seats to the highest-ranked remaining students, regardless of their type, who apply to $s^o$. 
	The result of this DA procedure determines the set of chosen students along with their seat types.
\end{itemize}
We call the corresponding DA mechanism \textbf{simultaneous-open-reserve} (SimOR).

$\C^{SimOR}$ is open-first verifiable. While there may exist other open-first verifiable choice rules, Lemma \ref{lemma:SimOR is open-first verifiable} proves that they must select the same set of students as $\C^{SimOR}$, and among them, $\C^{SimOR}$ always assigns the minimum number of reserved seats. For a choice rule $ \C $ and a set of applicants $ I $, we respectively use $ \C^r\langle I \rangle$ and $ \C^o\langle I \rangle$ to denote the set of chosen students who occupy reserved seats and the set of chosen students who occupy open seats.

\begin{lemma}\label{lemma:SimOR is open-first verifiable}
	$\C^{SimOR}$ is open-first verifiable. For any other open-first verifiable choice rule $\C$ and any $I\in 2^\I$, $\C\langle I \rangle=\C^{SimOR}\langle I \rangle$ and $ \C^{SimOR,r}\langle I \rangle \subseteq \C^r\langle I \rangle$.
\end{lemma}

Example \ref{example:open-first verifiable} illustrates the selection procedure of $\C^{SimOR}$ and its difference from other open-first verifiable choice rules.

\begin{example} \label{example:open-first verifiable}
	Consider a school $s$ that faces ten students $I=\{i_1,i_2,\ldots,i_{10}\}$. Students belong to three types, $I^{m_1} = \{i_1,i_2,i_3,i_4\}$, $I^{m_2} = \{i_5,i_6,i_7,i_8\}$, and $I^{m_3} = \{i_9,i_{10}\}$. School $s$ ranks the students as $i_1\succ_s i_2\succ_s i_3\succ_s i_5\succ_s i_4\succ_s i_9\succ_s i_6\succ_s i_7\succ_s i_8\succ_s i_{10}$.
	Suppose that school $ s $ has eight seats, with $q_s^o=3$, $q_s^{m_1}=2$, $q_s^{m_2}=1$, and $q_s^{m_3}=2$. 
	
	To generate $\C^{SimOR}(I)$, in the first step of the DA procedure, all students apply to $s^o$. Then, $s$ rejects $i_8 $ and $i_{10}$ and assigns open seats to the others. In the second step, $i_8 $ and $i_{10}$ apply to $s^r$. Then, $s$ assigns reserved seats to $i_8 $ and $i_{10}$ and rejects $i_6$ and $i_7$. In the last step, $i_6$ and $i_7$ apply to $s^r$. Then, $s$ assigns a reserved seat to $i_6$ and rejects $i_7$ and $i_8$. So, $$\C^{SimOR}(I)=\begin{pmatrix}
		s^o & s^o & s^o& s^o& s^o& s^o& s^r & s^r \\
		i_1 & i_2 & i_3 & i_5 & i_4 & i_9 & i_6 & i_{10}  
	\end{pmatrix}.
	$$
	
	There are two other open-first verifiable assignments for this example:
	\begin{center}
		$\begin{pmatrix}
			s^o & s^o & s^o& s^o& s^o& s^r& s^r & s^r \\
			i_1 & i_2 & i_3 & i_5 & i_4 & i_9 & i_6 & i_{10}  
		\end{pmatrix}
		$ and $\begin{pmatrix}
			s^o & s^o & s^o& s^o& s^r& s^r& s^r & s^r \\
			i_1 & i_2 & i_3 & i_5 & i_4 & i_9 & i_6 & i_{10}  
		\end{pmatrix}.
		$ 
	\end{center}
	These assignments select the same set of students but assign more reserved seats than $ \C^{SimOR} $.
\end{example}

The following theorem is our first main result in the paper.

\begin{theorem}\label{theorem:choicerule:characterization}
	
	A choice rule $\C$ is reserve-first verifiable if and only if it is $\C^{SimRO}$. A choice rule $\C$ is open-first verifiable if and only if, for all $I\in 2^\I$, $ \C\langle I \rangle=\C^{SimOR}\langle I \rangle$ and there exists $j\in \C^{SimOR,o}\langle I \rangle$ such that $\C^r\langle I \rangle=\{i\in \C^{SimOR}\langle I \rangle: j\succ_s i\}$.
\end{theorem}

In words, $\C$ is open-first verifiable if and only if it always selects the same set of students as $ \C^{SimOR} $, and there exists a threshold student $j$ who occupies an open seat under $ \C^{SimOR} $ such that all selected students ranked below $j$ are assigned reserved seats, and all selected students ranked at or above $j$ are assigned open seats. $\C^{SimOR}$ itself corresponds to choosing $j$ as the lowest-priority open-seat student, thereby minimizing reserved-seat assignments.

Let $C^{SimRO}$ and $C^{SimOR}$ denote the standard choice rule induced by $\C^{SimRO}$ and $\C^{SimOR}$, respectively. We obtain the following corollary.

\begin{corollary}\label{corollary:choicerule:characterization}
	A standard choice rule is induced by a reserve-first verifiable (resp.\ open-first verifiable) choice rule if and only if it is $C^{SimRO}$ (resp.\ $C^{SimOR}$).
\end{corollary}

\subsection{Characterization of Verifiable and Strategy-proof Mechanisms}\label{section:mechanism:characterization}

We extend our characterization of verifiable choice rules to a characterization of verifiable and strategy-proof mechanisms. 
By Lemma \ref{lemma:verifiability}, as long as schools use substitutable and verifiable choice rules, the corresponding DA mechanism is individually rational, strategy-proof, and verifiable. However, it does not preclude the potential existence of other mechanisms that possess these properties but do not belong to the class of DA mechanisms. In the following, we show that all such mechanisms must be essentially equivalent to those DA mechanisms.

We utilize the result of \cite{imamuraRecipe} to obtain our characterization. In a matching with contracts model, \cite{imamuraRecipe} call an axiom for standard choice rules \textit{punctual} if it imposes a requirement on choice problems that can be checked without referring to other choice problems. They show that, if every school uses a standard choice rule that is path independent, size monotonic, and characterized by a set of punctual axioms, then the corresponding DA is the unique mechanism satisfying individual rationality, strategy-proofness, and extensions of those punctual axioms to matchings. 

In our paper, verifiability restricts how a school selects students from a set of applicants. So, it is punctual. By ignoring seat types, Corollary \ref{corollary:choicerule:characterization} shows that reserve-first and open-first verifiability respectively characterize $C^{SimRO}$ and $C^{SimOR}$, both of which are path independent and size monotonic standard choice rules. The extensions of these axioms to matchings are essentially equivalent to our definition of verifiability for outcomes, in a sense made precise in Appendix \ref{appendix:section:simultaneous}. This allows us to derive Lemma \ref{lemma:mechanism:characterization} based on the main result of \cite{imamuraRecipe}. In Lemma \ref{lemma:mechanism:characterization}, a standard direct mechanism refers to a function $\varphi$ that finds a standard matching $\varphi(P_\I)$ for every preference profile $P_\I$. A standard direct mechanism $\varphi$ is induced by a direct mechanism $ \psi $ if, for every $P_\I$, $\varphi(P_\I)$ is a matching induced by the outcome $\psi(P_\I)$.  

\begin{lemma}\label{lemma:mechanism:characterization}
	A standard direct mechanism is individually rational, strategy-proof, and induced by a reserve-first verifiable (resp.\ open-first verifiable) direct mechanism if and only if it is induced by SimRO (resp.\ SimOR).
\end{lemma}

We apply Lemma \ref{lemma:mechanism:characterization} to prove Theorem \ref{theorem:mechanism:characterization}.

\begin{theorem}\label{theorem:mechanism:characterization}
	Let $ \psi $ be a direct mechanism that is individually rational, strategy-proof, and verifiable:	
	(1) If $ \psi $ is reserve-first verifiable, then $ \psi $ is equivalent to SimRO. 
	
	(2) If $ \psi $ is open-first verifiable, then for every preference profile $ P_\I $, students are assigned to the same schools under both $ \psi(P_\I) $ and $SimOR(P_\I) $, yet the set of students occupying reserved seats in $ \psi(P_\I) $ contains those in $SimOR(P_\I) $.
\end{theorem}

Theorem \ref{theorem:mechanism:characterization} shows that verifiability and strategy-proofness single out exactly SimRO and the class of mechanisms with school assignments equivalent to SimOR.

\section{Implications for Admission Clearinghouses in Practice}\label{section:applicationtoChina}

Having characterized the verifiable mechanisms, we now turn to practice. This section uses the verifiability lens to interpret how admission clearinghouses are actually designed across Chinese cities, to derive concrete guidance for policymakers who must choose among competing mechanisms, and to extend that guidance to other countries.

\subsection{How Admission Clearinghouses Are Designed in China}\label{section:practice}

Chinese policymakers have adopted a diverse set of mechanisms for high school admissions. \autoref{tab:mechanisms-in-china} compiles the mechanisms used by 35 major Chinese cities in 2023. 
Among them, 23 cities adopt sequential mechanisms and 12 adopt simultaneous mechanisms; among the 23, 18 use SeqRO and 5 use SeqOR. The potential wastefulness of SeqOR may explain why SeqRO predominates. Among simultaneous mechanisms, SimRO and SimORO are most commonly used; Section~\ref{section:choicerule:characterization} has shown that SimRO is verifiable while SimORO is not.

Verifiability organizes this variation. Some cities that initially used the unverifiable SimORO have since switched to SeqRO,\footnote{For example, Shenzhen used SimORO before 2022 and switched to SeqRO in 2022. See the linked document of \href{https://www.sz.gov.cn/cn/xxgk/zfxxgj/zwdt/content/post_9703422.html}{Shenzhen Municipal Education Bureau (2022)}.} a transition consistent with policymakers' pursuit of verifiability. A natural question is why these policymakers do not instead switch to the verifiable SimRO. The answer reveals a second design objective: many policymakers adopting simultaneous mechanisms prefer to fill open seats before reserved seats, so as to leave reserved seats for students from lower-performing middle schools, and some state this goal explicitly.\footnote{For example, the document of \href{https://jyj.hefei.gov.cn/jydt/tzgg/18846553.html}{Hefei Municipal Education Bureau (2025)} stated such a goal.} $\C^{SimOR}$ is the verifiable rule that best serves this goal, but it is not straightforward to implement: it is defined via DA in an artificial economy in which students rank open seats above reserved seats, which differs from how policymakers typically design and communicate admission procedures.

Both $\C^{SimRO}$ and $\C^{SimOR}$ impose a fixed priority between open and reserved seats. Section \ref{section:SimFlex} examines another approach observed in several Chinese cities that allows each student to choose her own seat-type priority order, making the seat-type assignment self-explanatory. We show that this approach provides a unified framework for $\C^{SimRO}$ and $\C^{SimOR}$, as they arise as the two extreme cases in which all students choose to prioritize open seats or reserved seats. Our construction of $\C^{SimOR}$ is in fact inspired by this approach.  Section~\ref{section:SimOR:case} then presents arguments for recommending $\C^{SimOR}$ among verifiable choice rules. Section \ref{section:other:countries} discusses applications to other countries.

\subsection{The Subschool Approach: Unifying $\C^{SimRO}$ and $\C^{SimOR}$}\label{section:SimFlex}

The key design principle of the subschool approach is to make seat-type determination transparent by giving each student control over her own seat-type priority order. The mechanism in this approach, we call \textbf{simultaneous-separate} (SimSep), explicitly divides each school $s$ into two subschools: the \textbf{open subschool} $s^o$, which allocates open seats, and the \textbf{reserved subschool} $s^r$, which allocates reserved seats. Students submit preferences over all subschools and DA is run to find an outcome. Several Chinese cities have adopted this approach precisely to help students understand how their seat types are determined.

We describe SimSep as a DA process in which each school uses a choice rule $\C^{SimSep}$. In each round, each unassigned student applies to the highest-ranked subschool in her preference list that has not rejected her. School $s$ receives a set of applicants $I$ applying to $s^o$ and a set $I'$ applying to $s^r$, represented by the pair $(I, I')$.

\begin{itemize}
	\item For every pair $(I, I')$, $\mathbf{\C^{SimSep}}$ selects students as follows: Subschool $s^o$ allocates open seats to the highest-ranked applicants from $I$, up to the quota $q^o_s$. Meanwhile, for each $m\in\M$, subschool $s^r$ allocates reserved seats to the highest-ranked applicants from $I'\cap\I^m$, up to the quota $q^m_s$. Let $\C^{SimSep}(I, I')$ denote the outcome and $\C^{SimSep}\langle I, I'\rangle$ the set of selected students.
\end{itemize}

An obvious limitation of SimSep is that its subschool quotas are fixed exogenously, so unfilled reserved seats may be wasted. A natural improvement is to allow unfilled reserved seats to be reallocated as open seats. We call this modified choice rule $\C^{SimFlex}$ and the corresponding DA mechanism \textbf{simultaneous-flexible} (SimFlex).

\begin{itemize}
	\item For every pair $(I, I')$, $\mathbf{\C^{SimFlex}}$ selects students as follows: In the first step, for each $m\in\M$, school $s$ allocates reserved seats to the highest-ranked students from $I'\cap\I^m$, up to the quota $q^m_s$. In the second step, school $s$ allocates all remaining seats as open seats to the highest-ranked students from $I$, regardless of type. Let $\C^{SimFlex}(I, I')$ denote the outcome and $\C^{SimFlex}\langle I, I'\rangle$ the set of selected students.
\end{itemize}

Since unfilled reserved seats are converted to open seats, $\C^{SimSep}\langle I, I'\rangle\subseteq \C^{SimFlex}\langle I, I'\rangle$ for every pair $(I, I')$. By Theorem~2 of \cite{chambers2017choice}, students are weakly better off under SimFlex than under SimSep.

Although students must submit preferences over subschools, they have no strict preference between $s^o$ and $s^r$ for any given school: they care about school assignment, not seat type. We show that it is a weakly dominant strategy for every student to truthfully rank subschools: if student $ i $ prefers $s$ over $\tilde{s}$, she ranks $ s^o $ and $ s^r $ above $ \tilde{s}^o $ and $ \tilde{s}^r $, implying that $ i $ must rank $ s^o $ and $ s^r $ adjacently in her preferences. Therefore, we say that a preference list over subschools $P^*_i$ is \textit{consistent} with student $i$'s preferences over schools $P_i$ if, for every school $s$, the subschools $s^o$ and $s^r$ are ranked adjacently and the relative school order matches $P_i$.

\begin{proposition}\label{proposition:SimFlex}
	In SimSep and SimFlex, it is a weakly dominant strategy for every student to submit a preference list consistent with her true preferences over schools. When all students play weakly dominant strategies, the outcome of SimFlex is stable.
\end{proposition}

For each school $s$, we say that student $i$ reports the order $s^o$--$s^r$ if she ranks $s^o$ immediately above $s^r$, and $s^r$--$s^o$ if she ranks $s^r$ immediately above $s^o$. Both orders are weakly dominant under SimFlex, and reporting either leads to the same school assignment for $i$, though her seat type may differ.

When all students play weakly dominant strategies under SimFlex, since every student ranks every pair of $s^o$ and $s^r$ adjacently, once a student is about to apply to any subschool of school $s$, we let $i$ apply to $s$ and let $\C^*$ be a choice rule that encapsulates the part of the SimFlex procedure governing how $s$ processes applications.
\begin{itemize}
	\item Facing applicants $I$ reporting $s^o$--$s^r$ and applicants $I'$ reporting $s^r$--$s^o$, $\mathbf{\C^*}$ selects students by running DA in an artificial economy consisting of $s$ and $I\cup I'$, where all $i\in I$ prefer $s^o$ over $s^r$ and all $j\in I'$ prefer $s^r$ over $s^o$, with $s$ using the choice rule $\C^{SimFlex}$. Denote the outcome by $\C^*(I, I')$.
\end{itemize}
Then, SimFlex is equivalent to a DA mechanism with the choice rule $\C^*$.

The following result is the main theoretical contribution of this section: the two verifiable choice rules characterized in Theorem~\ref{theorem:choicerule:characterization} arise as the two extreme cases of $\C^*$.

\begin{proposition}\label{proposition:unify}
	For all $I\in  2^{\I}$, $\C^{SimOR}(I)=\C^*(I,\emptyset)$ and $\C^{SimRO}(I)=\C^*(\emptyset,I)$.
\end{proposition}

When all students report $s^o$--$s^r$ (so $I'=\emptyset$), $\C^*$ implements $\C^{SimOR}$; when all report $s^r$--$s^o$ (so $I=\emptyset$), it implements $\C^{SimRO}$. The two characterized choice rules are thus not ad hoc constructions but the natural endpoints of a single flexible mechanism under different student coordination outcomes. Under either pure coordination outcome, SimFlex is verifiable: uniform reporting of $s^o$--$s^r$ yields an open-first verifiable outcome, and uniform reporting of $s^r$--$s^o$ yields a reserve-first verifiable outcome. When students mix reporting orders, the outcome may fail to be verifiable, as Example~\ref{example:C_SimFlex is not verifiable} in Appendix~\ref{appendix:section:examples} shows.

Among the 35 cities in \autoref{tab:mechanisms-in-china}, Shenyang and Xiamen adopt SimSep.\footnote{See the linked documents of \href{https://jyj.shenyang.gov.cn/zwgkzdgz/fdzdgknr/jyta/zxta/202208/t20220815_3941431.html}{Shenyang Municipal Education Bureau (2021)} and of \href{https://edu.xm.gov.cn/jyfw/zsks/zkzz/202507/t20250710_2944484.htm}{Xiamen Municipal Education Bureau (2025)}.\label{footnote:shenyang}} Xiamen explicitly requires students to rank $s^o$ above $s^r$ for every school; under weakly dominant strategies, Xiamen's mechanism therefore approximates SimOR. Shenyang imposes no such requirement. A government proposal noted that many middle schools preferred their students to list open seats above reserved seats and suggested adopting the same requirement as Xiamen; Shenyang's policymakers ultimately chose not to do so.\footnote{See the linked document of Shenyang in \autoref{footnote:shenyang}.}

Although reporting $s^o$--$s^r$ is not individually required in Shenyang, students may nonetheless have a collective incentive to do so. Since both $s^o$--$s^r$ and $s^r$--$s^o$ are weakly dominant, switching from the latter to the former is costless for student $i$. The following result shows that this switch weakly expands the set of $i$'s peers who are selected by school $s$.

\begin{proposition} \label{proposition:best collective strategy}
	For any pair of disjoint applicant sets $(I,I')$ and any $i\in I'$, $\C^*\langle I,I'\rangle \cap \I^{\tau(i)} \subseteq \C^*\langle I\cup \{i\},I'\backslash \{i\}\rangle \cap \I^{\tau(i)}$.
\end{proposition}

When all students switch to reporting $s^o$--$s^r$, by Proposition~\ref{proposition:unify}, the outcome converges to SimOR.\footnote{Proposition~\ref{proposition:best collective strategy} holds at the choice-rule level. Whether peers benefit at the outcome level depends on the full preference profile, but one would expect that statistically more peers benefit than suffer.} This provides a collective rationality foundation for the Xiamen requirement: even absent a mandate, each student has an incentive to report $s^o$--$s^r$ if she values the welfare of her peers. It also provides an explanation for the preferences of middle schools in Shenyang.

\subsection{Choosing a Mechanism: Guidance for Policymakers}\label{section:SimOR:case}

A policymaker who has committed to disclosing only cutoffs, and therefore to a verifiable mechanism, still faces a choice between $\C^{SimRO}$ and $\C^{SimOR}$. We recommend $\C^{SimOR}$, for at least three reasons that bear directly on practice.

\textit{Distributional consequences.} The primary argument concerns how $\C^{SimRO}$ and $\C^{SimOR}$ differ in their treatment of targeted students. Under $\C^{SimRO}$, reserved seats are allocated before open seats: a high-scoring targeted student who claims a reserved seat reduces the reserved seats available to lower-scoring peers who may need affirmative action more. Under $\C^{SimOR}$, open seats are allocated first, so reserved seats remain for targeted students who could not secure an open seat on merit. In the case of two student types, $\C^{SimOR}$ and $\C^{SimRO}$ coincide with the rules studied by \cite{dur2018reserve}. Their results imply that, within a broad class of choice rules, $\C^{SimOR}$ maximizes the number of targeted students admitted, whereas $\C^{SimRO}$ minimizes it. For multiple types, no single maximization criterion is available, but the preference for SimOR persists: it consistently leaves reserved seats for lower-scoring targeted students rather than absorbing them with higher-scoring peers. Many Chinese cities explicitly state the goal of reserving seats for lower-scoring students, and their preference for open-first mechanisms reflects this priority.

\textit{Transparency about affirmative action use.} Among open-first verifiable choice rules, $\C^{SimOR}$ minimizes reserved-seat assignments, as established in Lemma~\ref{lemma:SimOR is open-first verifiable}. This is a transparency property: $\C^{SimOR}$ admits as many students as possible on merit before deploying reserved seats, revealing the true extent to which affirmative action is actually used. If a city designates 50\% of seats as reserved but finds that only 30\% are claimed, the policy may appear less redistributive than initially perceived, enhancing its political sustainability. A mechanism that fills reserve quotas mechanically even when targeted students would have qualified for open seats inflates the apparent footprint of affirmative action and may generate unnecessary opposition.

\textit{Floor constraints.} In the context of China, $ \C^{SimOR} $ provides an additional benefit for targeted students. As mentioned in the related literature, several Chinese cities impose a floor constraint on reserved-seat admissions, requiring that the score of a reserved-seat student not fall below the lowest open-seat score at the same school by more than a fixed margin. Consequently, a low-scoring targeted student may be rejected by a school despite unfilled reserved seats. By maximizing open-seat admissions, $ \C^{SimOR} $ minimizes the open-seat cutoff at each school, thereby relaxing the floor constraint and broadening access to reserved seats for the students who most need them.

\subsection{Applications to Brazil and India}\label{section:other:countries}

The paper's framework extends beyond China to other countries with affirmative action in centralized admissions. Two prominent examples are Brazil and India, which operate large-scale centralized admissions under constitutional or legislative mandates for affirmative action. In both countries, the key elements of the paper's framework, including multiple types of reserved seats, category-specific cutoffs, and minimal public disclosure, are present. The verifiability concept and Theorem~\ref{theorem:choicerule:characterization} apply, while some institutional details require extensions (e.g., horizontal reserves in India).

Brazil's centralized university admission system allocates seats between a universal open category and multiple quota categories for disadvantaged groups, using a national exam score as the priority index. For each institution and program, the system publicly discloses the minimum qualifying score in each category, corresponding to the cutoff profile $c(\mu)$ in our framework. This public disclosure reflects a commitment to transparency: students and the public can verify whether the quota policy is correctly applied by comparing individual scores to the published cutoffs. The remaining question of whether the published cutoffs alone also determine which category's seat a student occupies is precisely what our verifiability concept addresses, and Theorem~\ref{theorem:choicerule:characterization} characterizes the allocation rules that guarantee it. The distributional arguments for $\C^{SimOR}$ in Section~\ref{section:SimOR:case} apply equally to the Brazilian context, since an open-first allocation leaves reserved seats for students from disadvantaged groups who could not secure an open seat on merit.

India's centralized university admissions and government recruitment systems allocate seats between an open general category and multiple reservation categories for historically disadvantaged groups, with publicly disclosed cutoffs for each combination of category, institution, and program. This multi-category structure with public cutoff disclosure fits the paper's multi-type framework directly, and the routine use of published cutoffs by students, families, and advocacy groups to audit whether the reservation policy is correctly implemented reflects the same verifiability goal. Theorem~\ref{theorem:choicerule:characterization} identifies the allocation rules under which seat-type assignment is verifiable from the published cutoffs alone, and the characterization applies for any number of reservation categories. One feature of India's system that goes beyond the current model is the distinction between ``vertical'' reservations and ``horizontal'' reservations. Incorporating horizontal reservations would require an extension of the framework, but the core seat-type verification question remains the same.

\section{Related Choice Rules in the Literature}\label{section:other:choicerule}

\subsection{Slot-specific Priorities (and Capacity Transfers)}\label{section:slot-specific}

We show that $\C^{SimRO}$ and $\C^{SimOR}$ do not belong to the class of choice rules with slot-specific priorities defined by \cite{kominers2016matching}. In this class, school $s$ has $q_s$ seats $\mathcal{A}=\{a_1,\ldots,a_{q_s}\}$, partitioned into open seats and reserved seats for each type via a function $\beta:\mathcal{A}\to\M\cup\{o\}$ with $|\beta^{-1}(o)|=q_s^o$ and $|\beta^{-1}(m)|=q_s^m$ for each $m\in\M$. A \emph{precedence order} $\rhd$ ranks the $q_s$ seats. The induced choice rule $\C^\rhd$ fills seats one by one in this order: an open seat $a_k^\rhd$ is assigned to the highest-priority remaining student under $\succ_s$; a reserved seat for type $m$ is assigned to the highest-priority remaining student under $\succ_m$, where students in $\I^m$ are ranked above all others while maintaining relative order within $\I^m$ and within $\I\backslash\I^m$. If the student admitted to a reserved seat does not belong to $\I^m$, the seat is reclassified as an open seat.

For any precedence order $\rhd$, $\C^\rhd$ is not verifiable. Moreover, $\C^\rhd$ may differ from $\C^{SimOR}$ and $\C^{SimRO}$ even when only the admitted students are compared; Examples~\ref{example:slot-specific and SimOR} and \ref{example:slot-specific and SimRO} show this. The fundamental difficulty with $\C^\rhd$ is that when a student type has unfilled reserve quotas, the precedence order determines which students fill them. But the set of eligible students varies across choice problems; it depends on which types have unfilled quotas. No fixed order can accommodate all such variations while maintaining verifiability.

\cite{avataneo2021slotspecific} extend this framework to slot-specific priorities with capacity transfers (SSPwCT), which subsume the dynamic reserve choice rules of \cite{aygun2020dynamic}. Each original seat has a shadow seat with initial capacity zero, and a capacity transfer vector governs transfers from unfilled original seats to their shadows. The choice rule $\C^{SimRO}$ belongs to the SSPwCT class. However, $\C^{SimOR}$ does not, even when only the admitted students are considered; this is illustrated by Example~\ref{example:SSPwCT}.

\subsection{Backward Transfer Choice Rule}\label{section:backward}

\cite{aygun2023how} introduce the \emph{backward transfer choice rule} to manage de-reservation of unfilled reserved seats for a specific category (the OBC) in India's affirmative action system, consistent with the over-and-above principle. The rule first allocates open seats to the highest-priority students, then allocates reserved seats to the highest-priority remaining students of each category. If reserved seats remain unfilled for OBC, their count is added to the open-seat quota and the first step is repeated; this continues until no OBC reserved seats remain unfilled or all applicants are admitted.

The choice rule $\C^{SimORO}$ is not verifiable because it does not ensure that students occupying open seats have higher priorities than those occupying reserved seats. Applying the backward transfer logic to $\C^{SimORO}$ yields an open-first verifiable choice rule: in each iteration, open seats are allocated to the highest-priority students regardless of type, reserved seats are allocated to the highest-priority remaining students of each type, and any unfilled reserved seats are transferred to the open quota. By Theorem~\ref{theorem:choicerule:characterization}, this procedure selects the same students as $\C^{SimOR}$. However, it may assign more students to reserved seats than $\C^{SimOR}$ does. For example, if all seats are filled after the first two steps of $\C^{SimORO}$, the backward transfer procedure leaves the outcome unchanged, whereas $\C^{SimOR}$ would assign all admitted students to open seats.

\subsection{Characterization Result of \cite{abdulkadiroglu2025market}}\label{section:comparison with regular rule}

\cite{abdulkadiroglu2025market} characterize the class of reserves-and-quotas choice rules in a school choice model with type-specific lower and upper bounds. Within this class, they recommend the \textit{regular rule}, which minimizes priority violations and selects the priority rank-maximal set of applicants.\footnote{The regular rule is also characterized by \cite{imamura2025meritocracy} through a different approach.} They argue that focusing on the regular rule is without loss of generality, since the outcome of any rule in the class can be replicated by the regular rule by simply increasing the lower bounds (reserve quotas).

In the absence of type-specific upper bounds, the regular rule coincides with $\C^{SimRO}$. The construction of $\C^{SimOR}$ is similar in spirit to that of the regular rule: both divide each school into subschools and run DA in an artificial economy.
The difference lies in how subschools are constructed and how students are assigned to subschools. \cite{abdulkadiroglu2025market} introduce type-specific subschools and allow students to apply to all subschools in any order. When all students first apply to their own type's subschool, the order of their remaining applications becomes irrelevant and the outcome is the regular rule. In our construction, motivated by real-world mechanisms used in China (Section \ref{section:SimFlex}), each school is divided only into two subschools. When all students first apply to the open subschool and then to the reserve subschool, the outcome is $\C^{SimOR}$. Example~\ref{example:SimOR with type maximal rule} shows that $\C^{SimOR}$ is neither priority violation-maximal nor priority rank-minimal within \citeauthor{abdulkadiroglu2025market}'s class of choice rules, so their characterization of the regular rule does not apply symmetrically to $\C^{SimOR}$.

Our different recommendations reflect distinct views on the purpose of affirmative action. We do not treat minimizing priority violations as a central objective. Instead, we view reserve quotas as a ``budget'' for priority violations: the goal is first to maximize the impact of affirmative action, and then, conditional on that, to minimize the expenditure of this budget. This leads to our recommendation of $\C^{SimOR}$. In practice, adjusting reserve quotas is often infeasible due to legal and political constraints, so we do not expect that the outcome of $\C^{SimOR}$ can be conveniently replicated by $\C^{SimRO}$ via increasing reserve quotas.

\section{Conclusion} \label{section:conclusion}

This paper studies how to design a centralized admission clearinghouse so that affirmative action is verifiable: each participant can confirm both her assigned school and her seat type from her own score and the publicly disclosed cutoffs. We show that verifiability, together with individual rationality and strategy-proofness, singles out essentially two deferred acceptance mechanisms, and we characterize the choice rules behind them. Among the two rules, $\C^{SimOR}$ is new and deploys a reserved seat only when a participant cannot earn an open seat on merit, so that every reserved-seat assignment reflects genuine affirmative action and the clearinghouse never overstates the policy it applies.

Beyond the characterization, the analysis yields guidance for the design of real admission clearinghouses. Verifiability organizes the diversity of mechanisms across 35 major Chinese cities. The same framework applies to the national admission systems of Brazil and India, where category-specific cutoffs are likewise disclosed, suggesting that verifiability is a practical design benchmark well beyond the Chinese context.

More broadly, our results show how public disclosure constraints can discipline the design of allocation systems. In markets where the policymaker controls both the allocation algorithm and the information released to participants, verifiability emerges as a first-order design requirement alongside incentive compatibility and fairness. Several extensions are natural and practically relevant: incorporating the horizontal reservations observed in some systems, accommodating richer disclosure regimes, and empirically evaluating how verifiability shapes participants' trust and behavior. We view these as promising directions for bringing transparency by design closer to the operation of large-scale matching markets.

\clearpage

\appendix

\section{Proofs}\label{appendix:proofs:bigsection}

\subsection{Proofs for Section \ref{section:verifiability:concept}} \label{appendix:sequential:proof}

\begin{proof}[\normalfont \textbf{Proof of Lemma \ref{lemma:stable}}]
\underline{``If'' part.} Individual rationality holds because $\emptyset$ is always achievable. Non-wastefulness holds because $|\mu(s)|<q_s$ implies $c^o_s=0$, making $s$ achievable to every student. Reserve-non-wastefulness holds because $sP_i\mu(i)$ implies $c^{\tau(i)}_s>0$, hence $q^{\tau(i)}_s=0$ or $|\mu^{r\tau(i)}(s)|=q^{\tau(i)}_s$. For no justified envy, suppose a student $i$ has justified envy towards another $j$, so $\pi_{i,\mu(j)}>\pi_{j,\mu(j)}$. Then $j$'s score satisfies $\pi_{j,\mu(j)}\ge c^o_{\mu(j)}$ if $t^\mu_j=o$, or $\pi_{j,\mu(j)}\ge\min\{c^{\tau(i)}_{\mu(j)},c^o_{\mu(j)}\}$ if $\tau(i)=\tau(j)$. Since $\pi_{i,\mu(j)}>\pi_{j,\mu(j)}$, in either case $\mu(j)$ is achievable to $i$, contradicting $\mu(j)P_i\mu(i)$.

\underline{``Only if'' part.} Suppose a student $i$ is not admitted to her best achievable school $s$, so $sP_i\mu(i)$. If $c^o_s=0$, then $|\mu(s)|<q_s$, so $\mu$ is wasteful. If $c^o_s>0$ and $c^{\tau(i)}_s=0$, then $q^{\tau(i)}_s>0$ and $|\mu^{r\tau(i)}(s)|<q^{\tau(i)}_s$, so $\mu$ violates reserve-non-wastefulness. If $c^o_s>0$ and $c^{\tau(i)}_s>0$, then $\pi_{i,s}>\min\{c^{\tau(i)}_s,c^o_s\}$, so $i$ has justified envy towards some $j\in\mu(s)$, contradicting stability of $\mu$.
\end{proof}

\subsection{Proofs for Section \ref{section:simultaneous}}\label{appendix:section:simultaneous}

\begin{proof}[\normalfont \textbf{Proof of Lemma \ref{lemma:verifiability}}]
Individual rationality is immediate since students never apply to unacceptable schools. Strategy-proofness and stability follow from \cite{hatfield2005matching}. For verifiability, fix any $P_\I$ and let $\mu$ be the DA outcome. Since DA coincides with the cumulative offer process under substitutability, $\C(\mu(s)\cup\{i\in\I:sP_i\mu(i)\})=\mu(s)$ for every school $s$. If $\C$ is reserve-first verifiable, then $\mu(s)$ is a reserve-first verifiable outcome in the economy consisting of $s$ and the students in $\mu(s)\cup\{i:sP_i\mu(i)\}$. So, the verifiability condition holds for every $i\in\mu(s)$, and $\mu$ is reserve-first verifiable. The open-first case is analogous.
\end{proof}

\begin{proof}[\normalfont \textbf{Proof of Lemma \ref{lemma:SimOR is open-first verifiable}}]
\underline{$\C^{SimOR}$ is open-first verifiable.} For any $I\in 2^\I$, the DA procedure generating $\C^{SimOR}(I)$ runs in the artificial economy consisting of $s$ and $I$ in which all students prefer $s^o$ over $s^r$. Proposition \ref{proposition:unify} in Section \ref{section:SimFlex} shows that the choice rule $\C^*$ unifies $\C^{SimOR}$ and $\C^{SimRO}$, and the stability of SimFlex under weakly dominant strategies proved in Proposition~\ref{proposition:SimFlex} implies that $\C^{SimOR}(I)$ is stable in that artificial economy. So, students' assignments can be characterized by school cutoffs. Since all students apply to $s^o$ before $s^r$ in the DA procedure, every student admitted to open seats has higher priority than every student admitted to reserved seats. So, $\C^{SimOR}(I)$ is open-first verifiable.

\underline{Any two open-first verifiable rules $\C$ and $\tilde{\C}$ satisfy $\C\langle I\rangle=\tilde{\C}\langle I\rangle$ for all $I\in 2^\I$.} Because verifiability implies stability, if $|I|\le q_s$, then $\C\langle I \rangle=\tilde{\C}\langle I \rangle =I$. For $|I|> q_s$, let $(c^o,\{c^m\}_{m\in \M})$ and $(\tilde{c}^o,\{\tilde{c}^m\}_{m\in \M})$ denote the cutoffs in $\C(I)$ and $\tilde{\C}(I)$.

If $\tilde{c}^o=c^o$, the same students occupy open seats in both outcomes; stability then forces the same students in reserved seats, giving $\C\langle I\rangle=\tilde{\C}\langle I\rangle$.

If $\tilde{c}^o\neq c^o$, assume without loss of generality that $\tilde{c}^o>c^o$. Every open-seat student in $\tilde{\C}(I)$ also occupies an open seat in $\C(I)$. It remains to show that every reserved-seat student $i$ in $\tilde{\C}(I)$ is admitted in $\C(I)$. 
Fix such $i$ with type $m=\tau(i)$. 

If some type-$m$ student is rejected in $\tilde{\C}(I)$, stability requires that all $q^m_s$ reserved seats for type $m$ be filled in $\tilde{\C}(I)$. So $\tilde{c}^m>0$, and there are exactly $q^{m}_s$ type-$m$ students with scores between $\tilde{c}^{m}$ and $\tilde{c}^o$, with one student's score being $\tilde{c}^{m}$. We claim $\tilde{c}^m\ge c^m$: if $c^m>\tilde{c}^m$, then since $\tilde{c}^o>c^o$, there are fewer than $q^m_s$ type-$m$ students with scores in $[c^m,c^o)$, so the reserved seats for type $m$ cannot be filled in $\C(I)$, a contradiction. So, every type-$m$ reserved-seat student in $\tilde{\C}(I)$ is admitted in $\C(I)$.

If all type-$m$ students are admitted in $\tilde{\C}(I)$, then at most $q^m_s$ of them have scores below $\tilde{c}^o$, hence at most $q^m_s$ have scores below $c^o$ (since $c^o<\tilde{c}^o$). If some $j\in I^m$ is rejected in $\C(I)$, reserve-non-wastefulness would require $q^m_s$ type-$m$ students with scores below $c^o$ to fill the reserved seats. But since $j$ is rejected, there are at most $q^m_s-1$ candidates, a contradiction.

Therefore, every student admitted in $\tilde{\C}(I)$ is admitted in $\C(I)$. So $\C\langle I\rangle=\tilde{\C}\langle I\rangle$. Since $\C^{SimOR}$ is open-first verifiable, $\C^{SimOR}\langle I\rangle=\C\langle I\rangle$.

\underline{For every open-first verifiable $\C$, $\C^{SimOR,r}\langle I\rangle\subseteq\C^r\langle I\rangle$ for all $I\in 2^\I$.} The case $|I|\le q_s$ is immediate since $\C^{SimOR,r}(I)=\emptyset$. For $|I|>q_s$, let $(\hat{c}^o,\{\hat{c}^m\}_{m\in \M})$ be the cutoffs in $\C^{SimOR}(I)$. We show $\hat{c}^o\le c^o$. Suppose $\hat{c}^o>c^o$. Let $i$ be the highest-score student occupying a reserved seat in $\C^{SimOR}(I)$, so $i$ occupies an open seat in $\C(I)$. 

If $I^{\tau(i)}\backslash\C^{SimOR}\langle I\rangle\neq\emptyset$, stability requires that all $q^{\tau(i)}_s$ reserved seats for type $\tau(i)$ be filled in $\C^{SimOR}(I)$, and since $\C^{SimOR}\langle I\rangle=\C\langle I\rangle$, those reserved seats are also filled in $\C(I)$. However, because $i$ occupies a reserved seat in $\C^{SimOR}(I)$ but an open seat in $\C(I)$, there are not enough lower-priority type-$\tau(i)$ students to fill the reserved seats in $\C(I)$, a contradiction. So $I^{\tau(i)}\backslash\C^{SimOR}\langle I\rangle=\emptyset$.

We now show that there exists $j\in I\backslash\C^{SimOR}\langle I\rangle$ with higher score than $i$, which means that $j\in I\backslash\C\langle I\rangle$ and $j$ has justified envy towards $i$ in $\C(I)$, a contradiction. To find $j$, since the DA procedure for generating $\C^{SimOR}(I)$ is independent of the order of students' proposals, modify the DA procedure by withholding $i$'s proposal after $i$ is rejected by $s^o$, letting the remaining students settle, then allowing $i$ to apply to $s^r$. Since $|I|>q_s$, by non-wastefulness, $i$'s admission to $s^r$ must displace some student $j$. Since $I^{\tau(i)}\backslash\C^{SimOR}\langle I\rangle=\emptyset$, $j$ is not a same-type student admitted to $s^r$. So, $j$ must have been admitted to  $s^o$ before $i$ applies to $s^r$,  and thus $j$ has higher score than $i$. Since $i$ has the highest score among reserved-seat students in $\C^{SimOR}(I)$, $j$ cannot be admitted to $s^r$. So, $j\in I\backslash\C^{SimOR}\langle I\rangle$.
\end{proof}

\begin{proof}[\normalfont \textbf{Proof of Theorem \ref{theorem:choicerule:characterization}}]
(1) \underline{$\C^{SimRO}$ is reserve-first verifiable.} For any $I\in 2^\I$, $\C^{SimRO}(I)$ assigns students to reserved seats before open seats, ensuring that within each type, reserved-seat students have higher priority than open-seat students, who in turn have higher priority than rejected students. By Proposition~\ref{proposition:SimFlex} and Proposition \ref{proposition:unify} in Section \ref{section:SimFlex}, which unifies $\C^{SimOR}$ and $\C^{SimRO}$, $\C^{SimRO}(I)$ is stable in the artificial economy consisting of $s$ and $I$, and thus can be characterized by school cutoffs. So $\C^{SimRO}$ is reserve-first verifiable. The proofs of Proposition~\ref{proposition:SimFlex} and Proposition \ref{proposition:unify} do not invoke Theorem \ref{theorem:choicerule:characterization}, so there is no circular dependency.

\underline{If $\C$ is reserve-first verifiable, then $\C=\C^{SimRO}$.} Reserve-first verifiability imposes, within each type $m$: (i) reserved-seat students have higher priority than open-seat students, who have higher priority than rejected students; (ii) if $|I^m|\le q^m_s$, all type-$m$ students must occupy reserved seats, since otherwise the reserved cutoff is zero, requiring all those students to occupy reserved seats, a contradiction; (iii) if $|I^m|> q^m_s$, the reserved seats for type $m$ must be filled, since otherwise we can obtain the same contradiction as in (ii). Together with q-acceptance implied by verifiability, these constraints uniquely determine $\C(I)=\C^{SimRO}(I)$.

(2) \underline{``If'' part.} Let $\C$ satisfy the stated condition. So for all $I\in 2^\I$, $\C\langle I\rangle=\C^{SimOR}\langle I\rangle$ and there exists $j\in\C^{SimOR,o}\langle I\rangle$ such that $\C^r\langle I\rangle=\{i\in\C\langle I\rangle:j\succ_s i\}$. This gives $\C^{SimOR,r}\langle I\rangle\subseteq\C^r\langle I\rangle$ and open-seat students have higher priority than reserved-seat students in $\C(I)$. It remains to show that $\C(I)$ is stable in the artificial economy consisting of $s$ and $I$.

Since $\C\langle I\rangle=\C^{SimOR}\langle I\rangle$ and $\C^{SimOR}(I)$ is stable in that artificial economy, $\C(I)$ is individually rational and non-wasteful. For reserve-non-wastefulness, if $I^m\backslash\C\langle I\rangle\neq\emptyset$, then we also have $I^m\backslash\C^{SimOR}\langle I\rangle\neq\emptyset$. So the $q^m_s$ reserved seats for type $m$ are filled in $\C^{SimOR}(I)$. Since $\C^{SimOR,r}\langle I\rangle\subseteq\C^r\langle I\rangle$, they are also filled in $\C(I)$. For no justified envy, consider any $j\in I\backslash\C\langle I\rangle$. All open-seat students in $\C(I)$ also occupy open seats in $\C^{SimOR}(I)$ (since $\C^{SimOR,r}\langle I\rangle\subseteq\C^r\langle I\rangle$), so they have higher priority than $j$. Since $\C\langle I\rangle=\C^{SimOR}\langle I\rangle$, the same students fill the reserved seats for $j$'s type in both $\C(I)$ and $\C^{SimOR}(I)$, and by stability of $\C^{SimOR}(I)$, each such student has higher priority than $j$.

\underline{``Only if'' part.} By Lemma~\ref{lemma:SimOR is open-first verifiable}, any open-first verifiable $\C$ satisfies $\C\langle I\rangle=\C^{SimOR}\langle I\rangle$ and $\C^{SimOR,r}\langle I\rangle\subseteq\C^r\langle I\rangle$ for all $I\in 2^\I$. Let $j$ be the lowest-priority open-seat student in $\C(I)$. Then $\C^r\langle I\rangle=\{i\in\C\langle I\rangle:j\succ_s i\}$ and $j\in\C^{SimOR,o}\langle I\rangle$.
\end{proof}

\begin{proof}[\normalfont \textbf{Proof of Lemma \ref{lemma:mechanism:characterization}}]
Following \cite{imamuraRecipe}, we define axioms on standard choice rules and then extend them to standard matchings.

We say that a standard choice rule $C$ satisfies \textit{RO-axiom} if $C(I)=C^{SimRO}(I)$ for all $I\in 2^\I$, and \textit{OR-axiom} if $C(I)=C^{SimOR}(I)$ for all $I\in 2^\I$. Both axioms are clearly punctual as called by \cite{imamuraRecipe}. By construction, a standard choice rule satisfying these axioms must be $ C^{SimRO} $ or $ C^{SimOR} $. Then, following \cite{imamuraRecipe}, given a preference profile $P_\I$, we say that a standard matching $u$ satisfies the \textit{extension of RO-axiom} if $u(s)=C^{SimRO}(\{i\in \I:sR_iu(i)\})$ for all school $s$, and the \textit{extension of OR-axiom} if $u(s)=C^{SimOR}(\{i\in \I:sR_iu(i)\})$ for all school $s$.

We claim that $u$ satisfies the extension of RO-axiom if and only if it is induced by a reserve-first verifiable outcome. For the ``if'' direction, suppose $u$ is induced by a reserve-first verifiable outcome $\mu$. Then for each $s$, $\mu(s)$ is a reserve-first verifiable outcome in the economy consisting of $s$ and $\{i\in \I:sR_iu(i)\}$. Since $\C^{SimRO}$ is the only reserve-first verifiable rule, $\mu(s)=\C^{SimRO}(\{i\in \I:sR_iu(i)\})$. So $u$ satisfies the extension of RO-axiom. For the ``only if'' direction, given that $u$ satisfies the extension of RO-axiom, construct an outcome $\mu$ with $\mu(s)=\C^{SimRO}(\{i:sR_iu(i)\})$ for all $s$. So $\mu$ induces $u$. Since $\C^{SimRO}$ is reserve-first verifiable, $\mu$ is reserve-first verifiable.  

The analogous claim holds for OR-axiom and open-first verifiable outcomes.

	Theorem 1 of \cite{imamuraRecipe} proves that, when every school uses a standard choice rule that satisfies path independence and size monotonicity and that is characterized by a set of punctual axioms, a standard direct mechanism satisfies individual rationality, strategy-proofness, and the extensions of these axioms if and only if it is the DA mechanism with the choice rule.

	In our paper, since $C^{SimRO}$ and $C^{SimOR}$ are substitutable and q-acceptant, they satisfy path independence and size monotonicity. They are characterized by RO-axiom and OR-axiom, respectively. Theorem~1 of \cite{imamuraRecipe} then implies that a standard direct mechanism satisfies individual rationality, strategy-proofness, and the extension of RO-axiom (resp.\ OR-axiom) if and only if it is DA with $C^{SimRO}$ (resp.\ $C^{SimOR}$). The equivalence among reserve-first (resp.\ open-first) verifiable outcomes gives Lemma~\ref{lemma:mechanism:characterization}.
\end{proof}

\begin{proof}[\normalfont \textbf{Proof of Theorem \ref{theorem:mechanism:characterization}}]
Let $\psi$ be individually rational, strategy-proof, and verifiable. For every preference profile $P_\I$, let $u$ denote the standard matching induced by $\psi(P_\I)$; for every school $s$, construct the economy consisting of $s$ and the students in $\{i\in\I:sR_iu(i)\}$.

If $\psi$ is reserve-first verifiable, by Lemma~\ref{lemma:mechanism:characterization}, $\psi(P_\I)$ and $SimRO(P_\I)$ induce the same $u$. Then $\psi_s(P_\I)$ and $SimRO_s(P_\I)$ are both reserve-first verifiable in the constructed economy and admit the same students. Since $\C^{SimRO}$ is the unique reserve-first verifiable choice rule, $\psi_s(P_\I)=\C^{SimRO}(\{i:sR_iu(i)\})=SimRO_s(P_\I)$ for every $s$ and $P_\I$. Therefore, $\psi=SimRO$.

If $\psi$ is open-first verifiable, by Lemma~\ref{lemma:mechanism:characterization}, $\psi(P_\I)$ and $SimOR(P_\I)$ induce the same $u$. Then $\psi_s(P_\I)$ and $SimOR_s(P_\I)$ are both open-first verifiable in the constructed economy and admit the same students. Because $ SimOR_s(P_\I)=\C^{SimOR} (\{i\in \I:s R_i u(i) \})$ for every $s$ and $\C^{SimOR}$ minimizes reserved-seat assignments among all open-first verifiable choice rules, every student occupying a reserved seat in $ SimOR(P_\I) $ also occupies a reserved seat in $\psi(P_\I)$.
\end{proof}

\subsection{Proofs for Section \ref{section:applicationtoChina}}

\begin{proof}[\normalfont \textbf{Proof of Proposition \ref{proposition:SimFlex}}]
We embed the model into the matching-with-contracts framework of \cite{hatfield2005matching}, with $s^o$ and $s^r$ as two contracts at school $s$. Since $\C^{SimFlex}$ is substitutable and q-acceptant, SimFlex is strategy-proof in the sense that if a preference order over subschools is viewed as a student's true preferences, the student cannot obtain a strictly better assignment by misreporting preferences. Fix any student $i$ and any two preference orders over subschools $P^*_i$ and $P^\diamond_i$. Treat each in turn as $i$'s true preference over subschools, and let $\mu^*$ and $\mu^\diamond$ denote the corresponding outcome of SimFlex. When $P^*_i$ is $i$'s true preference, we have $\mu^*(i)R^*_i\mu^\diamond(i)$, whereas when $P^\diamond_i$ is $i$'s true preference, we have $\mu^\diamond(i)R^\diamond_i\mu^*(i)$. If $P^*_i$ is consistent with $i$'s original preferences over schools $P_i$, then $\mu^*(i)R^*_i\mu^\diamond(i)$ means that $i$ is assigned to a weakly better school according to $P_i$ by reporting $P^*_i$ than by reporting $P^\diamond_i$. If both $P^*_i$ and $P^\diamond_i$ are consistent with $P_i$, then $i$ must be assigned to the same school in $\mu^*$ and $\mu^\diamond$. So, it is a weakly dominant strategy for $i$ to report any preference order consistent with $P_i$. The same argument holds for SimSep.

When all students play weakly dominant strategies, the outcome of SimFlex is stable: individual rationality holds since no student lists a subschool of an unacceptable school as acceptable; non-wastefulness holds since $\C^{SimFlex}$ is q-acceptant; reserve-non-wastefulness holds since any student rejected by a school has her type's reserved seats filled; no justified envy holds since both seat types are allocated based on priority ranking.
\end{proof}

\begin{proof}[\normalfont \textbf{Proof of Proposition \ref{proposition:unify}}]
	For any $I\in  2^{\I}$, in the artificial economy consisting of school $s$ and $I$, if all $i\in I$ reports $ s^o$--$s^r$, then the DA procedure defining $\C^*(I,\emptyset)$ becomes equivalent to the DA procedure defining $\C^{SimOR}(I)$. So,     
    $\C^{SimOR}(I)=\C^*(I,\emptyset)$. Similarly, if all $i\in I$ reports $ s^r$--$s^o$, then the DA procedure defining $\C^*(\emptyset,I)$ becomes equivalent to the procedure defining $\C^{SimRO}(I)$. So, $\C^{SimRO}(I)=\C^*(\emptyset,I)$.
\end{proof}

\begin{proof}[\normalfont \textbf{Proof of Proposition \ref{proposition:best collective strategy}}]
	If $i \notin \C^*\langle I,I'\rangle$, then $ i $ is rejected by both $ s^o $ and $ s^r $ in the DA procedure generating $\C^*(I,I')$. Then, if $i$ switches to $ s^o$--$s^r $, $ i $ must still be rejected, and it does not change the result of the DA procedure. So, $ \C^*\langle I,I'\rangle =  \C^*\langle I\cup \{i\},I'\backslash \{i\} \rangle$.

	If $i \in \C^*\langle I,I'\rangle$, that is, $(s,i,t) \in \C^*(I,I')$ for some $t\in \mathcal{T}$, we then consider two cases. 
	
	If $(s,i,o) \in \C^*(I,I')$, it means that when $i$ reports $s^r$--$s^o$, $i$ is first rejected by $s^r$ and then admitted to $ s^o $. Then, when $ i $ switches to $s^o$--$s^r$, $ i $ must be admitted to $ s^o $, and the other assignments remain unchanged. So, $ \C^*\langle I,I'\rangle =  \C^*\langle I\cup \{i\},I'\backslash \{i\} \rangle$.
	
	If $(s,i,r) \in \C^*(I,I')$, by Proposition \ref{proposition:SimFlex}, $ i $ remains admitted by switching to $s^o$--$s^r$. If $(s,i,r) \in  \C^*(I\cup \{i\},I'\backslash \{i\})$, it means that when reporting $s^r$--$s^o$, $i$ is admitted to $ s^r $, while when switching to $s^o$--$s^r$, $ i $ is first rejected by $s^o$ and then admitted to $ s^r $, leaving other assignments unchanged. So, $ \C^*\langle I,I'\rangle =  \C^*\langle I\cup \{i\},I'\backslash \{i\} \rangle$.
	If $(s,i,o) \in  \C^*(I\cup \{i\},I'\backslash \{i\})$, it means that when reporting $s^r$--$s^o$, $i$ is admitted to $ s^r $, while when switching to $s^o$--$s^r$, $ i $ is admitted to $s^o$. So, one additional reserved seat for type $\tau(i)$ becomes available for lower-priority same-type students, giving $\C^*\langle I,I'\rangle \cap \I^{\tau(i)} \subseteq \C^*\langle I\cup \{i\},I'\backslash \{i\} \rangle \cap \I^{\tau(i)}$.
\end{proof}

\section{Examples}\label{appendix:section:examples}

\begin{example}[SimFlex is not verifiable] \label{example:C_SimFlex is not verifiable}
		Consider a school $ s $ and six students $\I = \{i_1,i_2,i_3,i_4,i_5,i_6\}$. Students belong to two types, $\I^{m_1}=\{i_1,i_2,i_3\}$ and $\I^{m_2}=\{i_4,i_5,i_6\}$. They are ranked by $s$ as $i_1\succ_s i_2\succ_s i_3\succ_s i_4\succ_s i_5\succ_s i_6$. Suppose that $q^o_s=2$, $q^{m_1}_s=2$, and $q^{m_2}_s=1$.   
		
		Suppose that students submit the following preferences under SimFlex:
		$$ \begin{matrix}
			i_1 & i_2 & i_3 & i_4 & i_5 & i_6 \\ \hline
			s^o  & s^o & s^r & s^r & s^o & s^o \\
			s^r  & s^r & s^o & s^o & s^r & s^r
		\end{matrix} $$  
		
		
		So, $I=\{i_1,i_2,i_5,i_6\}$ is the set of applicants who report $ s^o$--$s^r $, and $I'=\{i_3,i_4\}$ is the set of applicants who report $ s^r$--$s^o $. In the first step of the DA procedure generating $ \C^*(I,I') $, $\{i_1,i_2,i_5,i_6\}$ apply to $s^o$ and $\{i_3,i_4\}$ apply to $s^r$; $s$ first assigns reserved seats to $i_3$ and $i_4$ and then open seats to $i_1$, $ i_2 $, and $i_5$, rejecting $i_6$. In the second step, $i_6$ applies to $s^r$ and is rejected. So, $ \C^*(I,I')=  \begin{pmatrix}
			s^o & s^o & s^r & s^r & s^o\\
			i_1 & i_2 & i_3 & i_4 & i_5  
		\end{pmatrix} $, which is not verifiable.
	\end{example}

\begin{example}[$\C^\rhd \neq \C^{SimOR}$]\label{example:slot-specific and SimOR}
	Consider a set of six students $I=\{i_1,i_2,\ldots,i_6\}$ who belong to three types: $I^{m_1}=\{i_1,i_2\}$, $I^{m_2}=\{i_3,i_4\}$, and $I^{m_3}=\{i_5,i_6\}$. A school $s$ has five seats, with $q_s^o=2$ and $q_s^{m_1}=q_s^{m_2}=q_s^{m_3}=1$. Let $a^o_1$ and $a^o_2$ denote the two open seats, and let $a^m$ denote the reserved seat for type $m$. Suppose that there exists a precedence order $\rhd$ such that $\C^\rhd\langle I \rangle=\C^{SimOR}\langle I \rangle$ for all possible priority rankings of the six students.
	
	We first show that both open seats must be placed before each reserved seat in $\rhd$. Consider the priority order
	\[
		i_3\succ_s i_1\succ_s i_4\succ_s i_5\succ_s i_6\succ_s i_2.
	\]
	Under $\C^{SimOR}$, the selected students are $\{i_1,i_2,i_3,i_4,i_5\}$; student $i_2$ is selected through the type-$m_1$ reserved seat, while $i_6$ is rejected. If $a^{m_1}$ were placed before one of the two open seats, then $a^{m_1}$ would be assigned to $i_1$ before both open seats had been filled. The remaining four seats would then be assigned to the four highest-priority students among $\{i_3,i_4,i_5,i_6,i_2\}$, so $i_6$ would be selected and $i_2$ rejected. This contradicts $\C^\rhd\langle I \rangle=\C^{SimOR}\langle I \rangle$. Therefore, both open seats must precede $a^{m_1}$. 
	
	By symmetric arguments, both open seats must precede $a^{m_2}$ and $a^{m_3}$ as well. Thus, the two open seats must be placed before all three reserved seats.

	Now consider the priority order
	\[
		i_1\succ_s i_2\succ_s i_3\succ_s i_5\succ_s i_6\succ_s i_4.
	\]
	Under $\C^{SimOR}$, the selected students are $\{i_1,i_2,i_3,i_4,i_5\}$, so $i_4$ is selected and $i_6$ is rejected. Since the first two seats in $\rhd$ are open, they are assigned to $i_1$ and $i_2$. To select $i_4$, the type-$m_1$ reserved seat, which is now unfilled by any type-$m_1$ student, must be placed before the type-$m_2$ reserved seat; otherwise the type-$m_2$ reserved seat is assigned to $i_3$, and then $i_4$ cannot be selected. Hence, to obtain $\C^\rhd\langle I \rangle=\C^{SimOR}\langle I \rangle$, we must have $a^{m_1}\rhd a^{m_2}$.
	
	Finally, consider the symmetric priority order
	\[
		i_3\succ_s i_4\succ_s i_1\succ_s i_5\succ_s i_6\succ_s i_2.
	\]
	The same argument implies $a^{m_2}\rhd a^{m_1}$, a contradiction. So, there does not exist a precedence order $\rhd$ such that $\C^\rhd\langle I \rangle=\C^{SimOR}\langle I \rangle$ for all possible priority orders.
\end{example}

\begin{example}[$\C^\rhd \neq \C^{SimRO}$]\label{example:slot-specific and SimRO}
	Consider a set of eleven students $I=\{i_1,i_2,\ldots,i_{11}\}$ who belong to four types: $I^{m_1}=\{i_1,i_2,i_3\}$, $I^{m_2}=\{i_4,i_5,i_6\}$, $I^{m_3}=\{i_7,i_8,i_9\}$, and $I^{m_4}=\{i_{10},i_{11}\}$. A school $s$ has seven seats, with $q_s^o=1$, $q_s^{m_1}=q_s^{m_2}=2$, and $q_s^{m_3}=q_s^{m_4}=1$. Let $a^o$ denote the open seat, and let $a^m$ denote a reserved seat for type $m$. Suppose that there exists a precedence order $\rhd$ such that $\C^\rhd$ and $\C^{SimRO}$ always select the same set of students from all possible applicant sets for all possible priority rankings.
	
	We first show that all reserved seats must be placed before the open seat in $\rhd$.
	
	Consider type $m_1$ and the priority order
	\[
	i_1\succ_s i_2 \succ_s  \cdots \succ_s i_3.
	\]
	Then, $i_1,i_2\in \C^{SimRO}\langle I \rangle$ and $i_3\notin \C^{SimRO}\langle I \rangle$. Under $\C^\rhd$, since there are two reserved seats for type $m_1$, if $ i_1$ or $i_2$ does not occupy such a reserved seat, $ i_3 $ must be admitted to such a reserved seat, contradicting $\C^\rhd\langle I \rangle= \C^{SimRO}\langle I \rangle$. Therefore, $ i_1$ and $i_2$ must both occupy reserved seats  in $ \C^\rhd(I) $. This requires that the two reserved seats for type $m_1$ must be placed before the open seat $a^o$ in $ \rhd $. 
	
	By symmetric arguments, every reserved seat must precede the open seat in $\rhd$.
	
	Now consider the priority order
	\[
		i_1\succ_s i_2\succ_s i_4\succ_s i_5\succ_s i_7\succ_s i_8\succ_s i_9\succ_s i_{10}\succ_s i_{11}\succ_s i_6\succ_s i_3.
	\]
	Let $I'=\{i_1,i_2,i_3,i_7,i_8,i_9,i_{10},i_{11}\}$. Then, $	\C^{SimRO}\langle I'\rangle=\{i_1,i_2,i_7,i_8,i_9,i_{10},i_{11}\}$, and $i_3$ is rejected. Since $I'$ contains no type-$m_2$ students, the two type-$m_2$ reserved seats are assigned as open seats under $\C^\rhd$. To obtain $\C^\rhd\langle I' \rangle= \C^{SimRO}\langle I' \rangle$, both type-$m_1$ reserved seats must be placed before both type-$m_2$ reserved seats; otherwise, either $i_1$ or $i_2$ would occupy a type-$m_2$ seat, leaving a type-$m_1$ reserved seat for $i_3$ (that is, $i_3\in \C^\rhd\langle I' \rangle$), a contradiction.
	
	Next, let $I''=\{i_4,i_5,i_6,i_7,i_8,i_9,i_{10},i_{11}\}$. Then, $\C^{SimRO}\langle I''\rangle=\{i_4,i_5,i_7,i_8,i_9,i_{10},i_{11}\}$,
	and $i_6$ is rejected. Since the two type-$m_1$ reserved seats must precede the two type-$m_2$ reserved seats and $I''$ contains no type-$m_1$ students, those two type-$m_1$ seats are assigned as open seats to $i_4$ and $i_5$. Then, a type-$m_2$ reserved seat remains available for $i_6$. So $i_6\in \C^\rhd\langle I''\rangle$, contradicting $\C^\rhd\langle I'' \rangle= \C^{SimRO}\langle I'' \rangle$. Hence there does not exist a precedence order $\rhd$ such that $\C^\rhd$ and $\C^{SimRO}$ always select the same set of students.
\end{example}

\begin{example}[$\C^{SimOR}\neq$ ~SSPwCT]\label{example:SSPwCT}
  
	Consider a set of seven students $I=\{i_1,i_2,\ldots,i_{7}\}$ who belong to three types: $I^{m_1}=\{i_1,i_2\}$, $I^{m_2}=\{i_3,i_4\}$, and $I^{m_3}=\{i_5,i_6,i_7\}$. A school $s$ has five seats, with $q_s^o=2$ and $q_s^{m_1}=q_s^{m_2}=q_s^{m_3}=1$. Let $a^o$ denote an open seat and $a^m$ the reserved seat for type $m$. So, the five seats are represented by $\{a^{o}_1,a^{o}_2,a^{m_1},a^{m_2},a^{m_3}\}$.
	
	To define $C^{SSPwCT}$, we introduce a shadow seat $e^m$ for each reserved seat $a^m$. Shadow seats for open seats would be redundant and thus are not introduced. School $s$ has two precedence orders: one over original seats, $\rhd^O$; the other over shadow seats, $\rhd^E$. A location vector $L$  specifies the position of each shadow seat relative to the original seats. Then, $(\rhd^O, \rhd^E, L) $ determine the exact precedence order of the original and shadow seats. A capacity transfer scheme $ t$ determines whether to transfer the capacity of an empty original seat to its shadow seat, which initially has no capacity.  
	
	Suppose that there exist $(\rhd^O, \rhd^E, L,t)$  such that an SSPwCT rule and $\C^{SimOR}$ always select the same set of students. Then, it is without loss of generality to assume that the two original open seats are placed before the three original reserved seats in $\rhd^O$, because the two open seats must be filled before the other seats in $ \C^{SimOR}$. 

	First, consider the priority order
	\[
		i_1\succ_s i_2\succ_s i_3\succ_s i_5\succ_s i_6\succ_s i_7\succ_s i_4.
	\]
	Under $\C^{SimOR}$, the selected students are $\{i_1,i_2,i_3,i_4,i_5\}$. Under SSPwCT, after the two original open seats are assigned to $i_1$ and $i_2$, the original type-$m_1$ reserved seat is empty. To reproduce $\C^{SimOR}\langle I \rangle$, the transferred capacity from $a^{m_1}$ must be used before $a^{m_2}$; otherwise $a^{m_2}$ admits $i_3$, and the transferred capacity is taken by a higher-priority type-$m_3$ student before $i_4$ can be admitted. Thus, the SSPwCT order must satisfy $a^{m_1}\rhd e^{m_1}\rhd a^{m_2}$.
	
	Now, consider the symmetric priority order
	\[
		i_3\succ_s i_4\succ_s i_1\succ_s i_5\succ_s i_6\succ_s i_7\succ_s i_2.
	\]
	The same argument implies $a^{m_2}\rhd e^{m_2}\rhd a^{m_1}$, a contradiction. So, there does not exist an SSPwCT rule that always selects the same set of students as $\C^{SimOR}$ does.
\end{example}

\begin{example}[$\C^{SimOR}$ is neither priority violation maximal nor priority rank minimal] \label{example:SimOR with type maximal rule}
	Consider eleven students who belong to three types: $I^{m_1} = \{i_1,i_2,i_3,i_4\}$, $I^{m_2} = \{i_5,i_6,i_7,i_8\}$, and $I^{m_3} = \{i_9,i_{10},i_{11}\}$. A school $s $ has nine seats, with $q_s^o=4$, $q_s^{m_1}=2$, $q_s^{m_2}=2$, and $q_s^{m_3}=1$. The school ranks the students as $i_1 \succ i_2\succ i_3 \succ  i_4\succ i_5 \succ i_9\succ i_6 \succ i_{10}\succ i_7\succ i_8 \succ  i_{11}$.

	It is easy to verify that,   $\C^{SimOR}\langle I \rangle=\{i_1,i_2,i_3,i_4,i_5,i_9,i_6,i_{10},i_7\}$.
	
	Now, we consider the rule $C'$ in the class of reserves-and-quotas rules, which lets all type-$m_3$ students apply to the open subschool before applying to their own type-specific subschool, and all other students apply to their own type-specific subschool before applying to the open subschool in the artificial DA procedure. 
	
	Then,
	$C'(I)= \{i_1,i_2,i_3,i_4,i_5,i_9,i_6,i_{10},i_{11}\}$, which creates more priority violations than $\C^{SimOR}\langle I \rangle$ and is priority dominated by $\C^{SimOR}\langle I \rangle$. 
\end{example}

\section{Nash Equilibrium Analysis of Sequential Mechanisms}\label{appendix:NE:sequential}

\begin{proposition}\label{proposition:NE of the sequential-reserve-open}
	Under complete information, (1) the set of NE outcomes of SeqRO equals the set of stable outcomes;
	(2) The set of non-wasteful NE outcomes of SeqOR is a subset of stable outcomes. However, SeqOR may have a wasteful NE outcome.
\end{proposition}

\begin{proof}[\normalfont \textbf{Proof of Proposition \ref{proposition:NE of the sequential-reserve-open}}]

Under both mechanisms, every NE outcome is individually rational, because listing an unacceptable school in one’s preference ranking is never optimal. Throughout the proof, we rely on the following \textit{DA monotonicity} property: removing a student from applicants in stage 1  weakly improves all remaining students' assignments in stage 1, and therefore the set of students who proceed to stage 2 can only weakly shrink.

\medskip
(\textbf{SeqRO}) Recall that stage~1 allocates reserved seats and stage~2 allocates open seats.

\underline{Every NE outcome $\mu$ is stable.}

\textit{Reserve-non-wastefulness.} Suppose there exists a pair $(i,s)$ such that $sP_i\mu(i)$ and $|\mu^{r\tau(i)}(s)|<q^{\tau(i)}_s$. Let $i$ deviate to reporting only $s$ (as acceptable) in stage~1. DA monotonicity implies the remaining students weakly improve, so $s$ still has unfilled reserved seats for type $\tau(i)$ and must admit $i$. Contradiction.

\textit{Non-wastefulness.} Suppose there exists a pair $(i,s)$ such that $|\mu(s)|<q_s$ and $sP_i\mu(i)$. Reserve-non-wastefulness implies the reserved seats for type $\tau(i)$ at $s$ are full, so $s$ has unfilled open seats in stage~2. Let $i$ deviate to reporting only $s$ in both stages. DA monotonicity implies $s$ still has unfilled open seats in stage~2 and must admit $i$. Contradiction.

\textit{No justified envy.} Suppose there is a pair $(i,j)$ such that $\pi_{i,\mu(j)}>\pi_{j,\mu(j)}$ and $\mu(j)P_i\mu(i)$.

If $t^\mu_j=o$, then $j$ is admitted in stage~2. Let $i$ deviate to reporting no acceptable school in stage~1 and only $\mu(j)$ in stage~2. By DA monotonicity, the applicant set in stage 2 weakly shrinks, so the admitted students at $\mu(j)$ in stage~2 have weakly lower priority than before. Since $i$ has higher priority than $j$, who was admitted, $i$ is admitted to $\mu(j)$. Contradiction.

If $t^\mu_j=r$, then $\tau(i)=\tau(j)$ and $j$ is admitted in stage~1. Let $i$ deviate to reporting only $\mu(j)$ in stage~1. By DA monotonicity, the admitted students at $\mu(j)$ in stage~1 have weakly lower priority than before. Since $i\succ_{\mu(j)} j$ and $j$ was admitted, $i$ is admitted. Contradiction.

\underline{Every stable outcome $\mu$ is a NE.} We construct a strategy profile as follows: for $t^\mu_i=r$, student $i$ reports only $\mu(i)$ in stage~1 and true preferences in stage~2; for $t^\mu_i=o$, she reports no acceptable school in stage~1 and only $\mu(i)$ in stage~2; for $\mu(i)=\emptyset$, she reports true preferences in both stages. It is easy to verify that this strategy profile produces $\mu$.

We show that no deviation is profitable. For any $i$ and any school $s$ with $sP_i\mu(i)$ (or every acceptable $s$ if $\mu(i)=\emptyset$), stability implies: (i) the reserved seats for type $\tau(i)$ at $s$ are occupied by same-type students with strictly higher priority, so no stage-1 strategy admits $i$ to $s$; (ii) consequently, no deviation changes the open-seat quota at $s$ in stage~2; (iii) all open seats at $s$ are occupied by students with strictly higher priority, so no stage-2 strategy admits $i$ to $s$.

\medskip
(\textbf{SeqOR}) Recall that stage~1 allocates open seats and stage~2 allocates reserved seats. The proof that every non-wasteful NE outcome $\mu$ is stable follows the same structure as SeqRO with stages interchanged.

\textit{Reserve-non-wastefulness.} Suppose there exists a pair $(i,s)$ such that $sP_i\mu(i)$ and $|\mu^{r\tau(i)}(s)|<q^{\tau(i)}_s$. Let $i$ deviate to reporting no acceptable school in stage~1 and only $s$ in stage~2. DA monotonicity implies the applicant set in stage 2 weakly shrinks, so $s$ still has unfilled reserved seats for type $\tau(i)$ and admits $i$. Contradiction.

\textit{No justified envy.} Suppose there is a pair $(i,j)$ such that $\pi_{i,\mu(j)}>\pi_{j,\mu(j)}$ and $\mu(j)P_i\mu(i)$.

If $t^\mu_j=o$, then $j$ is admitted in stage~1. Let $i$ deviate to reporting only $\mu(j)$ in stage~1. Since $i\succ_{\mu(j)} j$ and $j$ was admitted to the open seat, $i$ is admitted. Contradiction.

If $t^\mu_j=r$, then $\tau(i)=\tau(j)$ and $j$ is admitted in stage~2. Let $i$ deviate to reporting no acceptable school in stage~1 and only $\mu(j)$ in stage~2. DA monotonicity implies the applicant set in stage 2 weakly shrinks, so the admitted students at $\mu(j)$ in stage~2 have weakly lower priority than before. Since $i\succ_{\mu(j)} j$ and $j$ was admitted to the reserved seat, $i$ is admitted. Contradiction.
\end{proof}

\begin{example}[Wasteful NE outcome of SeqOR] \label{example:sequential:manipulation}
	Consider four students $ \I=\{i_1,i_2,i_3,i_4\} $ and two schools $ \{s_1,s_2 \}$. Students belong to two types, $ \I^{m_1}=\{i_1,i_2\} $ and $ \I^{m_2}=\{i_3,i_4\} $. They all prefer $ s_1 $ over $ s_2 $. Schools have equal quotas: $ q^o_{s_1}=q^o_{s_2}=1 $ and $ q^{m_1}_{s_1}=q^{m_1}_{s_2}=1 $. Both schools rank students as $ i_1\succ i_2\succ i_3 \succ i_4 $.

	The following strategies constitute a NE under SeqOR that leads to a wasteful outcome: $ i_2 $ reports no acceptable schools for stage 1 and true preferences for stage 2, and the other students report true preferences for both stages. In stage 1, $ i_1 $ and $ i_3 $ occupy an open seat at $ s_1 $ and $ s_2 $, respectively. In stage 2, $ i_2 $ occupies a reserved seat at $ s_1 $, and $ i_4 $ is unmatched. The reserved seat at $ s_2 $ for $ \I^{m_1} $ is wasted.
	
\end{example}

\begin{example}[A stable outcome that is not a NE outcome of SeqOR]  
	\label{example:multipleNE}
	Consider four students $\I=\{i_1,i_2,i_3,i_4\}$ and two schools $\S=\{s_1,s_2\}$. Students belong to three types, $ \I^{m_1}=\{i_1\} $, $ \I^{m_2}=\{i_2\} $, and $ \I^{m_3}=\{i_3,i_4\} $. Students $ i_1 $ and $ i_3 $ prefer $ s_2 $ over $ s_1 $, while $ i_2 $ and $ i_4 $ prefer $s_1$ over $s_2$. Schools have the following quotas: $ q^o_{s_1}=q^o_{s_2}=1 $ and $ q^{m_1}_{s_1}=q^{m_2}_{s_2}=1 $. Both schools rank students as $  i_3 \succ i_4 \succ i_1\succ i_2$.

	There are two stable outcomes, \( \mu \) and \( \tilde{\mu} \). In both, \( i_3 \) and \( i_4 \) are admitted to open seats at \( s_2 \) and \( s_1 \), respectively. The two outcomes differ in the assignments of \( i_1 \) and \( i_2 \). In \( \mu \), \( i_1 \) and \( i_2 \) are admitted to reserved seats at \( s_1 \) and \( s_2 \), respectively. In \( \tilde{\mu} \), they swap schools and occupy open seats. Moreover, \( \tilde{\mu} \) Pareto dominates \( \mu \).

    Under SeqOR, $\tilde{\mu}$ cannot be supported as a NE outcome. Since reserved seats are set aside when open seats are allocated in stage 1, the two open seats in stage 1 must be allocated to $ i_3 $ and $ i_4 $. So, $ i_1 $ and $ i_2 $ must occupy reserved seats in stage 2. It is a NE for all students to report true preferences for both stages, which leads to $ \mu $.
    
	In contrast, under SeqRO, it is a NE for all students to report true preferences for both stages, which leads to $\mu$. It is also a NE for all students to report no acceptable schools for stage 1 and true preferences for stage 2, which leads to $\tilde{\mu}$.
\end{example}

\clearpage

	\section{Admission Mechanisms used in Major Chinese Cities}\label{appendix:citylist}

	\autoref{tab:mechanisms-in-china} lists the mechanisms used by 35 major cities in China in 2023. These cities include the capital cities of 30 provincial-level divisions and 5 other cities ranked among the top twenty by population (i.e., Dongguan, Qingdao,  Shenzhen, Suzhou, Xiamen). The capital city of Tibet is not listed because it does not implement the affirmative action policy we study.
	
	\setcounter{table}{0} 
	
	\begin{table}[!htbp]
		\caption{Mechanisms used by major cities of China in 2023}
		\label{tab:mechanisms-in-china}
		\centering
			{\footnotesize
				\begin{threeparttable}
					\begin{tabular}{lll}
						\hline
						City & Reserve quota & Mechanism \\ \hline
						Beijing      & 50\%      & Sequential-Reserve-Open \\
						Shanghai     & 60\%--65\% & Sequential-Reserve-Open \\
						Chongqing    & 70\%      & Sequential-Reserve-Open \\
						Changchun    & 60\%--80\% & Sequential-Reserve-Open \\
						Jinan        & 60\%      & Sequential-Reserve-Open \\
						Shijiazhuang & 80\%      & Sequential-Reserve-Open \\
						Nanjing      & 50\%      & Sequential-Reserve-Open \\
						Hangzhou     & 40\%--60\% & Sequential-Reserve-Open \\
						Nanchang     & 70\%      & Sequential-Reserve-Open \\
						Changsha     & 60\%      & Sequential-Reserve-Open \\
						Wuhan        & 50\%      & Sequential-Reserve-Open \\
						Guangzhou    & 50\%      & Sequential-Reserve-Open \\
						Chengdu      & 50\%      & Sequential-Reserve-Open \\
						Xining       & 50\%      & Sequential-Reserve-Open \\
						Yinchuan     & 60\%      & Sequential-Reserve-Open \tnote{1} \\
						Urumqi       & 40\%      & Sequential-Reserve-Open \\
						Shenzhen     & 50\%      & Sequential-Reserve-Open \\
						Suzhou       & 70\%      & Sequential-Reserve-Open \\ \hline
						Harbin       & 60\%      & Sequential-Open-Reserve \\
						Fuzhou       & 55\%      & Sequential-Open-Reserve \\
						Haikou       & 50\%      & Sequential-Open-Reserve \\
						Kunming      & 70\%      & Sequential-Open-Reserve \\
						Hohhot       & 50\%      & Sequential-Open-Reserve \tnote{2} \\ \hline
						Guiyang      & 50\%      & Simultaneous-Reserve-Open \\
						Qingdao      & 65\%      & Simultaneous-Reserve-Open \\
						Tianjin      & 50\%      & Simultaneous-Reserve-Open \\ \hline
						Taiyuan      & 60\%      & Simultaneous-Open-Reserve-Open \\
						Zhengzhou    & 60\%      & Simultaneous-Open-Reserve-Open \\
						Hefei        & 85\%      & Simultaneous-Open-Reserve-Open \\
						Nanning      & 50\%      & Simultaneous-Open-Reserve-Open \\
						Xi'an        & 50\%      & Simultaneous-Open-Reserve-Open \\
						Lanzhou      & 60\%--75\% & Simultaneous-Open-Reserve-Open \\
						Dongguan     & 50\%      & Simultaneous-Open-Reserve-Open \\ \hline
						Shenyang     & 70\%      & Simultaneous-Separate \\
						Xiamen       & 55\%      & Simultaneous-Separate \\ \hline
					\end{tabular}
					
					\begin{tablenotes}[flushleft]
						\footnotesize 
						\item[1] Yinchuan, capital of Ningxia Province, employs a dynamic mechanism for the allocation of open seats.  
						\item[2] Hohhot, capital of Inner Mongolia Province, employs a dynamic mechanism similar to its college admission process \citep{gong2025dynamic} in both stages. 
					\end{tablenotes}
				\end{threeparttable}
			}
	\end{table}

\clearpage
	
	\setlength{\bibsep}{0pt plus 0.2ex}
	\bibliographystyle{aea}
	\bibliography{reference,schoolchoice2}
	
\end{document}